\documentclass[english,french,british,review, authoryear]{elsarticle}
\usepackage[T1]{fontenc}
\usepackage[utf8x]{inputenc}
\usepackage{color}
\usepackage{babel}
\addto\extrasfrench{%
   \providecommand{\fg}{\ifdim\lastskip>\z@\unskip\fi~\frqq}%
}

\usepackage{textcomp}
\usepackage{url}
\usepackage{amsmath}
\usepackage{amssymb}
\usepackage{graphicx}
\usepackage[unicode=true,
 bookmarks=true,bookmarksnumbered=false,bookmarksopen=false,
 breaklinks=true,pdfborder={0 0 0},backref=false,colorlinks=true]
 {hyperref}
\hypersetup{
 pdfauthor={Sylvain Douté}}

\makeatletter

\DeclareFontEncoding{LGR}{}{}
\DeclareTextSymbol{\~}{LGR}{126}
\providecommand{\tabularnewline}{\\}

\newcommand\micron{\mbox{$\mu$m}}%
\newcommand{\ls}{L\ensuremath{_{S}}}%
\newcommand{\co}{CO\ensuremath{_{2}}}%
\newcommand{\ho}{H\ensuremath{_{2}}O}%

\let\jnl@style=\rm
\def\ref@jnl#1{{\jnl@style#1}}

\def\aj{\ref@jnl{AJ}}                   
\def\actaa{\ref@jnl{Acta Astron.}}      
\def\araa{\ref@jnl{ARA\&A}}             
\def\apj{\ref@jnl{ApJ}}                 
\def\apjl{\ref@jnl{ApJ}}                
\def\apjs{\ref@jnl{ApJS}}               
\def\ao{\ref@jnl{Appl.~Opt.}}           
\def\apss{\ref@jnl{Ap\&SS}}             
\def\aap{\ref@jnl{A\&A}}                
\def\aapr{\ref@jnl{A\&A~Rev.}}          
\def\aaps{\ref@jnl{A\&AS}}              
\def\azh{\ref@jnl{AZh}}                 
\def\baas{\ref@jnl{BAAS}}               
\def\bac{\ref@jnl{Bull. astr. Inst. Czechosl.}}
\def\caa{\ref@jnl{Chinese Astron. Astrophys.}}
\def\cjaa{\ref@jnl{Chinese J. Astron. Astrophys.}}
\def\icarus{\ref@jnl{Icarus}}           
\def\jcap{\ref@jnl{J. Cosmology Astropart. Phys.}}
\def\jrasc{\ref@jnl{JRASC}}             
\def\memras{\ref@jnl{MmRAS}}            
\def\mnras{\ref@jnl{MNRAS}}             
\def\na{\ref@jnl{New A}}                
\def\nar{\ref@jnl{New A Rev.}}          
\def\pra{\ref@jnl{Phys.~Rev.~A}}        
\def\prb{\ref@jnl{Phys.~Rev.~B}}        
\def\prc{\ref@jnl{Phys.~Rev.~C}}        
\def\prd{\ref@jnl{Phys.~Rev.~D}}        
\def\pre{\ref@jnl{Phys.~Rev.~E}}        
\def\prl{\ref@jnl{Phys.~Rev.~Lett.}}    
\def\pasa{\ref@jnl{PASA}}               
\def\pasp{\ref@jnl{PASP}}               
\def\pasj{\ref@jnl{PASJ}}               
\def\rmxaa{\ref@jnl{Rev. Mexicana Astron. Astrofis.}}%
\def\qjras{\ref@jnl{QJRAS}}             
\def\skytel{\ref@jnl{S\&T}}             
\def\solphys{\ref@jnl{Sol.~Phys.}}      
\def\sovast{\ref@jnl{Soviet~Ast.}}      
\def\ssr{\ref@jnl{Space~Sci.~Rev.}}     
\def\zap{\ref@jnl{ZAp}}                 
\def\nat{\ref@jnl{Nature}}              
\def\iaucirc{\ref@jnl{IAU~Circ.}}       
\def\aplett{\ref@jnl{Astrophys.~Lett.}} 
\def\apspr{\ref@jnl{Astrophys.~Space~Phys.~Res.}}
\def\bain{\ref@jnl{Bull.~Astron.~Inst.~Netherlands}} 
\def\fcp{\ref@jnl{Fund.~Cosmic~Phys.}}  
\def\gca{\ref@jnl{Geochim.~Cosmochim.~Acta}}   
\def\grl{\ref@jnl{Geophys.~Res.~Lett.}} 
\def\jcp{\ref@jnl{J.~Chem.~Phys.}}      
\def\jgr{\ref@jnl{J.~Geophys.~Res.}}    
\def\jqsrt{\ref@jnl{J.~Quant.~Spec.~Radiat.~Transf.}}
\def\memsai{\ref@jnl{Mem.~Soc.~Astron.~Italiana}}
\def\nphysa{\ref@jnl{Nucl.~Phys.~A}}   
\def\physrep{\ref@jnl{Phys.~Rep.}}   
\def\physscr{\ref@jnl{Phys.~Scr}}   
\def\planss{\ref@jnl{Planet.~Space~Sci.}}   
\def\procspie{\ref@jnl{Proc.~SPIE}}   

\makeatother

\begin{document}

\title{Monitoring Atmospheric Dust Spring Activity at High Southern Latitudes
on Mars using OMEGA}

\author{S. Douté$^{1}$}

\address{$^{1}$Institut de Planétologie et d'Astrophysique de Grenoble (IPAG),
France (sylvain.doute@obs.ujf-grenoble.fr Phone: +33 4 76 51 41 71
Fax +33 4 76 51 41 46)}
\begin{abstract}
This article presents a monitoring of the atmospheric dust in the
south polar region during spring of martian year 27. Our goal is to
contribute to identifying the  regions where the dust concentration
in the atmosphere shows specific temporal patterns, for instance high,
variable, and on the rise due to lifting or transport mechanisms.
This identification is performed in relation with the seasonal ice
regression. Based on a phenomenological examination of the previous
results, hypothesis regarding the origin of aerosol activity of the
southern polar region are proposed. This is of paramount importance
since local dust storms generated in this region sometimes grow to
global proportions. The imaging spectrometer OMEGA on board Mars Express
has acquired the most comprehensive set of observations to date in
the near-infrared (0.93-5.1 microns) of the southern high latitudes
of Mars from mid-winter solstice (Ls=110\textdegree{}, December 2004)
to the end of the recession at Ls=320\textdegree{} (November 2005).
We use two complementary methods in order to retrieve the optical
depth of the atmospheric dust at a reference wavelength of one micron.
The methods are independently operated for pixels showing mineral
surfaces on the one hand and the seasonal cap on the other hand. They
are applied on a time series of OMEGA images acquired between \ls=220\textdegree{}
and \ls=280\textdegree{}  . As a result the aerosol optical depth
(AOD) is mapped and binned at a spatial resolution of 1.0\textdegree{}.pixel$^{\text{-1}}$
and with a mean period of AOD sampling ranging from less than two
sols for latitudes higher than 80\textdegree{}S to approximately six
sols at latitudes in the interval 65-75\textdegree{}S. We then generate
and interpret time series of orthographic mosaics depicting the spatio-temporal
distribution of the seasonal mean values, the variance and the local
time dependence of the AOD. In particular we suspect  that two mechanisms
play a major role for lifting and transporting efficiently mineral
particles and create dust events or storms: (i) nighttime katabatic
winds at locations where a favourable combination of frozen terrains
and topography exists (ii) large scale ($\approx$ 10-100 km) daytime
 thermal circulations at the edge of the cap when the defrosting area
is sufficiently narrow. As regards to the source regions around the
cap, the  sector with the highest AOD values/variability/increase
spans longitudes 180-300\textdegree{}E around \ls $\approx$ 250\textdegree{}.
Later (\ls $\approx$ 267\textdegree{}) the cryptic sector becomes
the most productive while the longitude sector 300-60\textdegree{}E
remain moderately dust-generative.  Our work calls for new simulations
of the martian surface-atmosphere dynamics at mesoscales to reproduce
the observations and confirm the interpretations.\end{abstract}
\begin{keyword}
Mars; South Pole; Atmosphere; Dust; OMEGA.
\end{keyword}
\maketitle

\section*{Introduction}

The southern high latitudes of Mars are of great interest in spring
and summer because of their role in the dust cycle. Local dust storms
generated in this region sometimes develop to global storms, and a
prominent dust collar encircles the polar cap. Several experiments
aboard orbiters have recently contributed to elaborate and refine
this picture. 

The TES and MOC instruments of Mars Global Surveyor have provided
a regular, but Sun-synchronous, record of dust activity in the south
polar region.

\citet{Toigo2002,Imamura2011} produced global maps of dust distribution
by integrating TES individual 9 \textmu{}m optical depth measurements
averaged over a 5\textdegree{}-Ls period (respectively 10\textdegree{}-Ls)
and binned in 5x5\textdegree{} boxes (respectively 5x10\textdegree{},
latitude, longitude). Both sets of maps depict very distinct space
and time patterns of activity around the polar cap edge for the first
common Martian Year (MY) 24. These seasonal trends are sometimes in
contradiction. For the following Martian year 25 and 26, \citet{Imamura2011}
reported a great stability of the dust opacity disturbance compared
to MY 24. Thermal mapping of dust by TES was originally limited to
regions where the temperature and thus the emitted signal are sufficiently
high, thus precluding the monitoring of the seasonal polar cap itself.
Nevertheless \citet{Horne2009} modified the standard TES aerosol
retrieval algorithm to retrieve atmospheric dust and ice optical depth
values for each daytime spectrum in the TES database with a surface
temperature below 210 K. As a result maps of the seasonal and spatial
variation of dust and water ice optical depth activity over both poles
are presented, averaged over a 2\textdegree{}-Ls period and binned
in 2x2\textdegree{} boxes from late MY24 to early MY27. For the southern
high latitudes the greatest observed dust activity each year takes
place above the growing seasonal cap from late summer to the beginning
of winter. At other seasons dust opacity is in general much lower
but some interannual variability, e.g. the beginnings of MY 25 global
storm, blurs this pattern. 

Following the early work of \citet{James2001} that already noted
correlation of storm event locations with the receding southern polar
cap, Color MOC wide angle images were mosaicked together by \citet{Toigo2002}
to produce daily global maps. Such snapshots show very dynamic dust
activity near the edge of the retreating south seasonal ice cap throughout
mid and late southern spring, then a decline going to midsummer. Visible
MOC snapshots are limited in time coverage and do not provide quantitative
values of dust opacity. 

The imaging spectrometer OMEGA aboard Mars Express allowed to overcome
some limitations of the previous experiments since it acquired a comprehensive
set of global observations in the near-infrared (0.93-5.1 microns)
of the southern high latitudes of Mars in spring and summer. A detailed
study of the contribution of water ice aerosols to the OMEGA dataset
is provided by \citet{Langevin2007}. This study is based on the water
ice absorption bands at 1.5, 2, and 3 μm. In 2005 (MY 27) from mid-spring
to mid summer most OMEGA observations are nearly free of water ice
either as aerosols or on the surface of the southern seasonal cap.
\citet{Vincendon2008} performed the mapping of the optical depth
of dust aerosols above areas of the south polar cap constituted of
pure \co\ ice as a function of Ls for dates when the contribution
of water ice aerosols can be neglected. The average trend of the temporal
evolution is a low optical depth between Ls = 180\textdegree{} and
Ls = 250\textdegree{} ($\tau$$ $(2.6 μm) = 0.1–0.2), an increase
of atmospheric dust activity observed between Ls = 250\textdegree{}
and Ls = 270\textdegree{} ($\tau$(2.6 μm) = 0.3–0.6), and then a
decrease up to Ls = 310\textdegree{}. \citet{Vincendon2008} observed
rapid time variations which are specific to a given location in conjunction
to large spatial variations of the optical depth observed over scales
of a few tens of kilometres. 

Monitoring of dust activity in the high southern latitudes by the
previous experiments was accompanied by an important effort in modelling
and simulation in order to interpret the observations in terms of
processes. The results of General Circulation Models (GCM) suggest
that non convective wind stress lifting produces the peak in the atmospheric
dust opacity during southern spring and summer and that convective
(dust devil) lifting is responsible for the background opacity during
other seasons \citep{Basu2004,Kahre2006}. However the coarse spatial
resolution achieved by GCM limits our understanding, fostering specific
simulations conducted at mesoscales. The main picture that emerges
from the latter studies is that flows capable of lifting dust from
the surface can be achieved by a variety of conditions, the most likely
being cap edge thermal contrasts \citep{Toigo2002} but also topography
\citep{Siili1999}. Regional or synoptic baroclinic instabilities
as well as vertical convection in the boundary layer could also play
a role \citep{Imamura2011}. These conditions as well as dust loading
itself in the atmosphere can interfere constructively or destructively.

The previous compilation of observations and simulations show that
some uncertainties and opened questions remain regarding dust activity
in the high southern latitudes. First \citet{Toigo2002,Imamura2011}
indicate different area where the mean atmospheric dust loading is
well above background levels for the same MY 24. Such discrepancy
entails an uncertainty on the location of the main source regions
as a function of time. Second the relative importance  of the expected
mechanisms for dust lifting at local and regional scales has not yet
been clearly established. Third, dust activity around and inside the
seasonal cap has  only been reconstructed conjointly by \citet{Horne2009}
although with a very coarse spatial resolution and only as a mean
of cross-validating their two retrieval techniques. A more spatially
detailed and integrated monitoring could be of paramount importance
to investigate the atmospheric dynamics across the cap edge. Finally
the main frequency of dust cloud generation and the time they take
to dissipate are also apparently inconsistent when examining the results
of \citet{Toigo2002} and \citet{Imamura2011}: daily as opposed to
every 10-20 sols; a few hours as opposed to 10 sols. Could that be
reconciled?

In this paper we bring some new insights about dust activity in the
southern polar region by monitoring the dust both inside and around
the seasonal cap based on the OMEGA dataset acquired during MY 27.
At the same time, special attention is paid to the exact characteristics
of the cap edge based on the work of \citet{Schmidt2009}. The mapping
of the optical depth of atmospheric dust in the near infrared above
mineral surfaces is made possible by the development of a new method
that is proposed in \citet{Doute2013a} and that is shortly described
in Section \ref{sec:Methods}. Allied to the complementary method
by \citet{Vincendon2008}, it is applied to analyse the time series
of OMEGA observations thus producing hundred of opacity maps. The
latter are integrated into a common geographical grid and processed
by a special data procedure so as to generate a time series of mosaics.
The mosaics depict the seasonal dust loading as well as the day-to-day
variability and local time dependence of the dust optical depth according
to solar longitude (Section \ref{sec:Analysis-of-dataset}). The mosaics
are fully described and examined in Section \ref{sec:Trends}. As
a result a synthetic view of dust activity in the south polar atmosphere
in mid spring to early summer is established and discussed in Section
\ref{sec:Discussion}. Finally, in Section \ref{sec:Summary}, the
main points of our study are summarised.

\section{Methods for retrieving the optical depth\label{sec:Methods}}

\subsection{Above ice free surfaces}

In summary (see \citet{Doute2013a} for more details) the first method
that we operate is based on a parametrisation bringing in the mean
effective optical path length of photons through the atmosphere composed
of particles and gas. The effective path length determines, with local
altimetry and the meteorological situation, the absorption band depth
of gaseous \co. In the following we assume that the top-of-atmosphere
(TOA) reflectance factor $R^{k}$ measured by OMEGA is:

\selectlanguage{english}%
\[
R^{k}(\theta_{i},\theta_{e},\phi_{e})\approx T_{gaz}^{k}(h,lat,long)^{\epsilon(\theta_{i},\theta_{e},\phi_{e},\tau_{aer}^{k0},H_{scale},A_{surf}^{k})}R_{surf+aer}^{k}(\theta_{i},\theta_{e},\phi_{e})
\]
\foreignlanguage{british}{where the quantity $R_{surf+aer}^{k}$ is
the reflectance factor that would be measured in the absence of atmospheric
gases. The reflectance factor is defined as the ratio of the radiance
coming from the planet by the radiance that would come from a idealised
lambertian surface observed under the same geometrical conditions
(illumination and viewing). The parametrisation can be expressed as
follows. The gas contribute to the signal as a simple multiplicative
transmission filter which is the aerosol free vertical transmission
$T_{gaz}^{k}(h,lat,long)$ scaled by the mean effective optical path
length $\epsilon$. The transmission $T_{gaz}^{k}$ is calculated
ab-initio using a Line-By-Line radiative transfer model fed by the
compositional and thermal profiles given by the European Mars Climate
Database \citep{Forget1999,Forget2006b} (MY24 dust Scenario, solar
averaged conditions, no perturbations added to the mean values) for
a given date, location and altitude of Mars. Exponent $\epsilon$
depends on (i) acquisition geometry ($\theta_{i}$, $\theta_{e}$
,$\phi_{e}$ )(ii) type, abundance, and vertical scale height $H_{scale}$
of the particles (iii) surface Lambertian albedo $A_{surf}^{k}$.
All the previous parameters are assumed to be known from ancillary
data or previous studies, except the dust aerosol integrated abundance
and surface albedo. Information about the first quantity can be reduced
to one value of aerosol optical depth (AOD) at a reference wavelength
of 1\micron\ $\tau_{aer}^{k0}$ (channel $k_{0}$) if the intrinsic
optical properties of the particles are known. }

\selectlanguage{british}%
Factor $\epsilon$ can be further decomposed into two terms such that: 

\[
\epsilon(\theta_{i},\theta_{e},\phi_{e},\tau_{aer}^{k0},H_{scale},A_{surf}^{k})=\psi(\nu)\beta(\theta_{i},\theta_{e},\phi_{e},\tau_{aer}^{k0},H_{scale},A_{surf}^{k})
\]

On the one hand factor $\psi(\nu)$ allows a quick and simplified
calculation of the free gaseous transmission along the geometrical
pathlength\foreignlanguage{english}{ of acquisition $\nu=\frac{1}{\cos(\theta_{i})}+\frac{1}{\cos(\theta_{e})}$}
knowing the vertical transmission $T_{gaz}^{k}$. On the other hand
$\beta(\theta_{i},\theta_{e},\phi_{e},\tau_{aer}^{k0},H_{scale},A_{surf}^{k})$
is a precious new observable that expresses how the aerosols influence
the pathlength. It can be tabulated by performing radiative transfer
theoretical calculations or experimentally estimated for each spectro-pixel
of an OMEGA image. When factor $\beta$ is tabulated, the single scattering
albedo, optical depth spectral shape, and phase function retrieved
in the near-infrared by \citet{Vincendon2008} are used since these
properties are relevant for the phase angle range spanned by the data
set of nadir OMEGA observations that we consider. In addition after
a series of tests described in \citet{Doute2013a}, the dust scale
height $H_{scale}$ is fixed at a value of 11km in agreement with
\citet{Vincendon2008}. Experimental estimation of factor $\beta$
for a given pixel means evaluating the intensity of the 2 \micron\
absorption band of gaseous \co. Practically this is only possible
for surfaces spectrally dominated by minerals or water ice, even though
the procedure can be extended for spectra showing saturated 2 \micron\ \co\ ice
absorption band but with some remaining radiance coming from the surface
such as in the outer part of the seasonal cap \citep{Doute2013a}.
For that purpose we first need the observed spectrum and, secondly,
the corresponding transmission spectrum $T_{gaz}^{k}(h,lat,long)^{\psi(\nu)}$
computed ab-initio. Combining the estimation of $\beta$ with the
reflectance factor deep into the 2\micron\ band (channel $k_{1}$),
we get, by iterative inversion of the tabulated function $\beta(\theta_{i},\theta_{e},\phi_{e},\tau_{aer}^{k0},H_{scale},A_{surf}^{k_{1}})$,
the AOD $\tau_{aer}^{k0}$ and the surface Lambertian albedo $A_{surf}^{k1}$.
Validation of the proposed method shows that it is reliable if two
conditions are fulfilled: (i) the observation conditions provide large
incidence or/and emergence angles (ii) the aerosols are vertically
well mixed in the atmosphere. As for the first condition, experiments
conducted on OMEGA nadir looking observations with incidence angles
higher than $\approx$ 60-65\textdegree{} produce very satisfactory
results. As for the second condition, analysis of test synthetic data
demonstrates that dust must reach an altitude of $\approx$ 5-8 km
in order to be satisfactorily detected.

\subsection{Above surfaces constituted by pure \co\ ice.}

The second method by \citet{Vincendon2008}  is restricted to area
where \co\ deposits are not contaminated by dust and water, i.e.
above most places of the seasonal cap except the cryptic sector and
close to the sublimation front where sub-pixel spatial mixing of ice-covered
and ice-free surfaces is observed. The mapping is based on the assumption
that the reflectance in the 2.64 \micron\ saturated absorption band
of the surface \co\ ice is mainly due to the light scattered by
aerosols. The atmospheric \co\ (respectively \ho) gas absorption
at 2.7 (respectively 2.6 μm) has negligible impact. In this case the
reflectance factor varies monotonically as a function of the optical
depth for a given set of photometric angles. Therefore, the optical
depth can be unambiguously determined by comparing the observed reflectance
factor at 2.64 μm with a reference look-up table. A method for selecting
pixels free of dust contamination has been derived from the relationship
between the observed reflectance factor at 1.08 μm and the optical
depth modelled from the reflectance at 2.6 μm. Correlation of low
frequency spatial variations of optical depth with altitude can be
modelled with a well-mixed dust atmospheric component with a scale
height of 11 km.

\subsection{Complementarity and possible limitations of the methods}

As a conclusion the two methods are complementary since our approach
specifically treats  ice free areas (mineral surfaces), areas dominated
by \ho\ ice or areas where remains \co\  ice provided that the latter
shows specific spectral properties. Conversely the method of \citet{Vincendon2008}
is restricted to areas entirely covered by \co\ deposits with the
supplementary condition that the latter are not contaminated by dust
and water. The common strength of both methods relies in the ability
to provide estimation of the AOD for each pixel of a single image,
i.e. at a fixed geometry. As regards the uncertainties, limitations
on the knowledge of the optical properties of aerosols will induce
a possible, systematic, and uniform bias in the maps. For the first
method, empirical tests have shown that this bias is likely small
since the mean effective optical path length is moderately dependent
on the single scattering albedo and the phase function of the aerosols
in the range of models proposed by authors in the last few decades
\citep{Korablev2005}. For the second method, the overall absolute
level of the optical depth could be biased by a factor up to 1.36.
For both methods the assumption of a lambertian surface has little
effect on the results. Otherwise the relative uncertainty linked with
stochastic errors in the measurements or in the modelling is of the
order of 10\%.

\section{Analysis of a time series of OMEGA observations\label{sec:Analysis-of-dataset}}

\subsection{The dataset}

The imaging spectrometer OMEGA on board Mars Express has acquired
a comprehensive set of observations in the near-infrared (0.93-5.1
microns) in the southern high latitudes of Mars from mid-winter solstice
(Ls=110\textdegree{}, December 2004) to the end of the recession 
(Ls=320\textdegree{}, November 2005) of martian year MY=27. These
observations provide a global coverage of the region with a time resolution
ranging from 3 days to one month and a spatial resolution ranging
from 700 m to 10 km.pixel$^{\text{-1}}$. We refer to \citet{Langevin2007}
for a complete description. We have systematically processed a subset
of 284 observations from \ls =220\textdegree{} to 280\textdegree{}
by using the two complementary methods of Section \ref{sec:Methods}.
As a result, we obtain a series of corresponding $\tau_{aer}^{k0}$
maps in the image space (optical depth $\tau_{aer}$ of the atmospheric
dust at a reference wavelength of one micron) that were normalised
according to a reference altitude of 0 km and a scale height of $H_{scale}=11$
km in order to correct for changes due solely to varying atmospheric
height because of topography:

\[
\tau_{det}^{k0}=\tau_{aer}^{k0}\exp\left(h/H_{scale}\right)
\]

The timescale between two maps that partially overlap is frequently
between 0.5\textdegree{} and 1\textdegree{} of Ls.

\subsection{Specific definitions related to the recession of the south seasonal
deposits\label{sub:definitions-SSPC}}

In the following a special attention is paid to the possible relationships
between atmospheric dust activity and the seasonal ice regression
in spring. The latter phenomenon has been monitored from orbit for
decades. Different concepts, now being recognised widely, were introduced
to ease its description. Since we use these concepts in our own study
it is useful to give at this point some definitions. The first are
relative to the passage of the sublimation front that can be detected
in one given location in the visible, the infrared, or the near-infrared
ranges (see \citet{Schmidt2009} for a summary). The time evolution
of three physical quantities - respectively the albedo, the temperature,
and the strength of a \co\  ice diagnostic absorption band, all related
to the surface - is parametrised by an analytical model. The ``crocus
date'' coincides with the inflection of the parametric curve, supposed
to coincide with \co\ disappearance. The ``crocus line'' is the
set of locations, where the crocus dates are equal to a given solar
longitude Ls. Some studies also suggested that the edge of the seasonal
cap can be more precisely described as a transition zone where patches
of \co\  ice and dust coexist geographically over a certain spatial
extent. In \citet{Schmidt2009} the transition zone is characterised
using \co\  detection by the OMEGA instrument in the near infrared.
The ``outer (respectively inner) crocus line'' is defined as the
set of locations that contain for a given date a \co\  ice coverage
$\approx$ 1\% (respectively $\approx$ 99\%). The crocus lines thus
describe the local structure of the edge of \co\  frost deposits
that are regressing toward the high latitudes during spring and beginning
of summer. The next two definitions pertain to the interior of the
south seasonal cap in spring. Using TES data, \citet{Kieffer2000}
define the ``cryptic region'' where \co\  ice (at low temperature)
has a low albedo and also recesses faster. The ``cryptic sector''
occupies longitudes between 60\textdegree{}E and 230\textdegree{}E
whereas the ``anti-cryptic sector'' is the complementary (longitudes
between 140\textdegree{}W and 60\textdegree{}E). The ``anti-cryptic
sector'' notably contains the permanent cap.

\subsection{Integrating the AOD maps into a common geographical grid\label{sub:map_Integration}}

\begin{figure}
\includegraphics[width=1\textwidth]{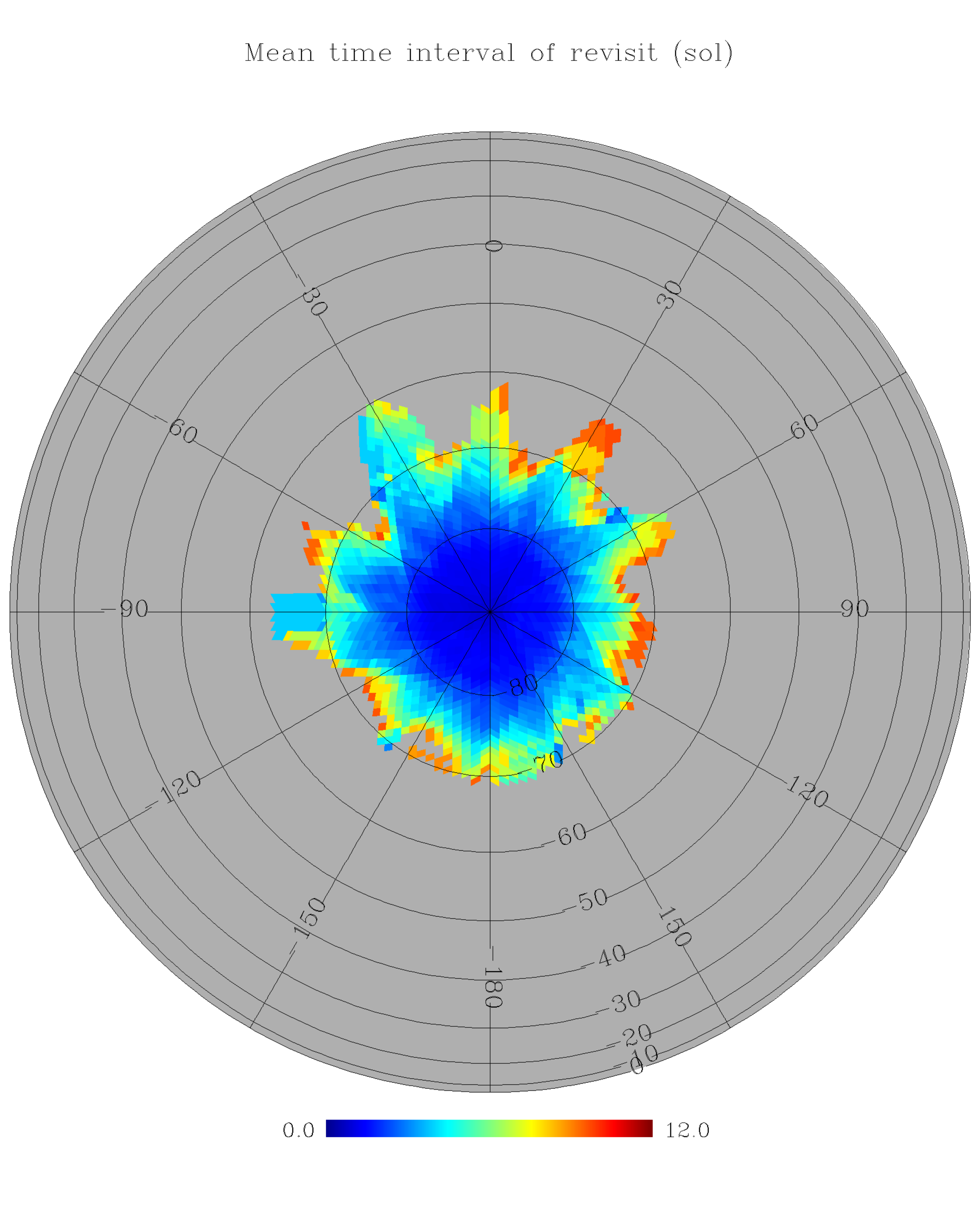}

\caption{Mean period (in sols) of aerosol optical depth sampling for a given
bin by the time series of OMEGA observations integrated into a common
geographical HEALPix grid of resolution 1.0\textdegree{}.pixel$^{\text{-1}}$.\label{fig:Mean-period-map}}
\end{figure}

These maps were independently integrated onto a common geographical
grid generated from the Hierarchical Equal Area isoLatitude Pixelization
(HEALPix, \url{http://healpix.jpl.nasa.gov}, \citep{Gorski2005})
of Mars southern hemisphere at different spatial resolutions. Such
an integration makes it easy to create a mosaic at a given date or
to build a time evolution curve at a given location. The resolution
of the grid is expressed by the parameter $N_{side}$ which defines
the number of divisions along the side of a base-resolution bin that
is needed to reach a desired high-resolution partition. The total
number of bins equal to $N_{pix}=12\times N_{side}^{2}$.Two built-in
properties of HEALPix - equal areas of discrete elements of partition,
and Iso-Latitude distribution of discrete area elements on the sphere
- make it easy to map any point of coordinate $(lat,long)$ into the
corresponding bin. The latter can also be tagged and addressed by
a single integer. If one considers in addition a partition of time
according to discrete solar longitudes - those of the observations
- any space-time data can be conveniently stored into a two dimensional
array (the Primary Integrated Data Array, PIDA). Its X dimension corresponds
to the bin number and its Y dimension corresponds to the time index.
Consequently the $\tau_{det}^{k0}$ maps - each corresponding to a
given image and thus date - are integrated, one at a time, by mapping
all the pixels where the AOD evaluation has succeeded to the appropriate
line of the array. In case several pixels fall into the same bin,
their values are averaged. After completion of the operation, the
whole collection of maps has been integrated into a common geographical
grid providing a mean period of AOD sampling for each bin that depends
basically on the latitude. As illustrated by Figure \ref{fig:Mean-period-map}
in the case $N_{side}$=64 (1.0\textdegree{}.pixel$^{\text{-1}}$)
the mean period ranges from less than two sols for latitudes higher
than 80\textdegree{}S to approximately six sols at latitudes in the
interval 65-75\textdegree{}S. An additional dimension can be optionally
added to the PIDA by considering the local time of acquisition for
each pixel. Then a division of the martian sol into three equal Local
Time (LT) intervals is adopted: 0-6, 6-12, and 12-18. The interval
18-24 is abandoned since it is only moderately populated by the OMEGA
spatio-temporal points of acquisition.

\subsection{Describing time evolutions\label{sub:describing-time-evolutions}}

\begin{figure}
\includegraphics[bb=10bp 150bp 550bp 600bp,clip,width=1\textwidth]{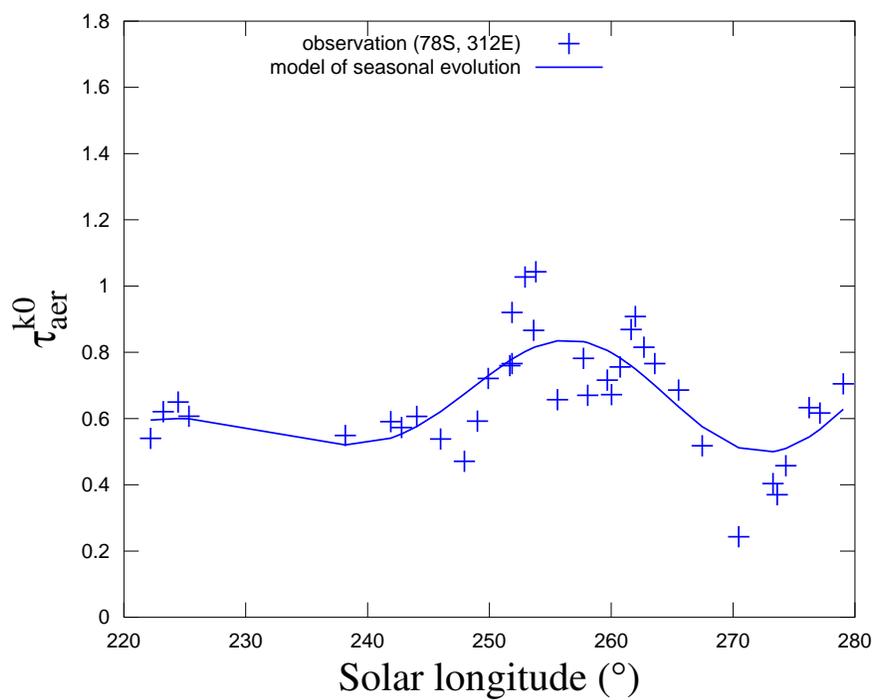}

\caption{Time evolution of the AOD for a given bin of the PIDA (crosses) and
modelling of its baseline by $\epsilon$-SVR (plain line).\label{fig:time_evolution_example}}
\end{figure}

Figure \ref{fig:time_evolution_example} shows as an example the time
evolution of the AOD for a bin chosen in the anticryptic longitude
sector (see definition in Section \ref{sub:definitions-SSPC}). It
was plotted by extracting a column of the PIDA. A noticeable day to
day variability (one data point to the next) can be immediately noted.
This can be expected for a very dynamic environment such as the polar
atmosphere in spring and beginning of summer. Variability could also
be accounted for by random errors that affect the AOD retrieval but,
with a theoretical root mean square of the order of $\approx0.05$,
they can only explain a part of it. A gap between \ls=228\textdegree{}
and \ls=235\textdegree{} and irregularities in the sampling are also
evident in the plot. Indeed coverage of the area by the OMEGA sensor
had necessary some limitations due to orbital characteristics, planning
constraints, and episodic OMEGA malfunctions. To mitigate the difficulties
induced by the previous factors on the analysis, we consider that
any of these temporal signals consists of two contributions: a mean
trend of $\tau_{det}^{k0}$ versus time, i.e. the baseline, and a
highly variable component, i.e. the variability around the baseline.
The former can be calculated by regression provided enough data points
are available. The latter is just the difference between the original
signal and the mean trend. Having a model for the seasonal trend in
the form of a regression function allows to fill the gaps in the PIDA
that is then restricted to the baseline component (Modified Integrated
Data Array, MIDA). Nevertheless this is done at the expense of the
spatial coverage as explained below. We now define the baseline of
any time evolution curve as the curvilinear object that minimises
its root mean square distance with the data points  (continuous line
in Figure \ref{fig:time_evolution_example}). The typical local curvature
of the baseline is controlled by a characteristic time scale that
is fixed at $\Delta L_{s}$=5\textdegree{}. Such parameter acts as
a threshold that separates what we consider to be a seasonal trend
 from what is the day-to-day variability. 

In order to calculate the baseline, we use Support Vector Machine
(SVMs) \citet{Vapnik1998} a popular machine learning method for classification,
regression, and other learning tasks. Traditional polynomial fit is
not suited to model the mean trend with its typical ups and downs
which will require using high orders. In addition SVM-based regression
allows us to control the characteristic time scale of the modelling.
Consider a set of data points, $\left\{ \left(\mathbf{x}_{1},y_{1}\right),\ldots,\left(\mathbf{x}_{l},y_{l}\right)\right\} $
where $x_{i}\in\mathbb{R}^{n}$ is a feature vector, $y_{i}\in\mathbb{R}^{1}$
is the target output, and $l$ is the number of  points available.
Under two given parameters $C>0$ and $\epsilon>0$ and the choice
of a kernel function $K$, the standard form of the Support Vector
Regression ($\epsilon$-SVR) is:

\[
\begin{array}{cl}
\underset{\mathbf{\alpha},\mathbf{\alpha}^{*}}{\min} & \frac{1}{2}\left(\mathbf{\alpha}-\mathbf{\alpha}^{*}\right)^{T}\mathbf{Q}\left(\mathbf{\alpha}-\mathbf{\alpha}^{*}\right)+\epsilon\underset{i=1}{\overset{l}{\sum}}\left(\alpha_{i}+\alpha_{i}^{*}\right)+\underset{i=1}{\overset{l}{\sum}}y_{i}\left(\alpha_{i}-\alpha_{i}^{*}\right)\\
\mathrm{subject\:\mathrm{to}} & \mathbf{e}^{T}\left(\mathbf{\alpha}-\mathbf{\alpha}^{*}\right)=0,\\
 & 0\leq\alpha_{i},\alpha_{i}^{*}\leq C,\: i=1,\cdots,l,
\end{array}
\]

where $Q_{ij}=K(\mathbf{x}_{i},\mathbf{x}_{j})$, and $\mathbf{e}=[1,\ldots,1]^{T}$
is the vector of all ones$ $. The solutions of the previous optimisation
problem is expressed in the form of $l$ support vectors $\alpha_{i}-\alpha_{i}^{*}$
and a constant $b$ such that the $\mathbb{R}$-valued approximate
regression function is:

\[
f(\mathbf{x})=\underset{i=1}{\overset{l}{\sum}}\left(-\alpha_{i}+\alpha_{i}^{*}\right)K(\mathbf{x}_{i},\mathbf{x})+b
\]

The self-adaptation of $\epsilon$-SVR to any kind of curvilinear
object is a decisive advantage in our case. For implementation of
the regression we use the library of support vector machine routines
LIBSVM \citep{Chang2011}, one of the most widely acclaimed SVM package.
We choose a radial basis function: $K(\mathbf{x}_{i},\mathbf{x}_{j})=\exp\left(-\gamma\left\Vert \mathbf{x}_{i}-\mathbf{x}_{j}\right\Vert ^{2}\right)$
where the parameter $\gamma$ controls the width of the function and
is directly related to the characteristic time scale $\gamma=2\Delta L_{s}$.
Regarding our regression, it should be noted that the feature vectors
reduce to scalars (solar longitudes) that must be linearly mapped
into $\left[0,1\right]$ following a requirement of the $\epsilon$-SVR
algorithm. The boundaries of the previous interval correspond respectively
to the minimum and maximum solar longitude considered in our study.
The parameter $\epsilon$ appears in the cost function and accommodates
the dispersion of the data. In our case it is fixed at 0.05. Finally
parameter $C$ is a regularisation term that is usually set to 1.0
by default. The LIBSVM routine for $\epsilon$-SVR directly outputs
\foreignlanguage{french}{$\mathbf{\alpha}-\mathbf{\alpha}^{*}$} and
$b$ allowing us to build the model of the mean trend of $\tau_{det}^{k0}$
versus time for any bin of the HEALPix grid for which enough points
are available and sufficiently distributed in the period of interest.
The bins for which these criteria are satisfied fall predominantly
poleward of the 70$^{\text{th}}$ parallel, the precise number depending
only slightly on the chosen spatial resolution of the grid. We find
that $N_{side}$=64 is the best compromise between the spatial resolution
and the number of curves to model.

\subsection{Generating time series of mosaics\label{sub:mosaics}}

Facilities associated to HEALPix provide a means to represent each
line of the PIDA or MIDA on a geographical map according to different
projections. In particular the orthographic projection is the most
suited among those available for the representation of the southern
polar region of Mars. Prior to the mapping of the MIDA, we average
along the Y dimension of the array all the valid values falling in
a given solar longitude interval (of width 10\textdegree{} for the
first two mosaics and then 2\textdegree{} for the following): \ls=220-230\textdegree{},
230-240\textdegree{}, 240-242\textdegree{}, and so on. In addition,
subsequently to the mapping, we superpose on the map the position
of the Seasonal South Polar Cap (SSPC) crocus lines as determined
by \citet{Schmidt2009} at the beginning of each time interval of
interest. As a result we obtain a series of 22 mosaics (at a spatial
resolution of $N_{side}$=64, i.e. 1.0\textdegree{}.pixel$^{\text{-1}}$)
that compiles the modelled version of the observations. Discussion
of the results will be principally based on  these mosaics. Map projections
of the individual lines of the PIDA lead to one AOD map at a spatial
resolution of $N_{side}$=1024, i.e. 1/16\textdegree{}.pixel$^{\text{-1}}$,
for each OMEGA observation. A selection of these maps (Figures \ref{fig:event0}
to \ref{fig:event4}) is marginally examined to get some hints about
the lowest latitudes and around the cap edge.

\subsection{Classifying the seasonal trends of $\tau_{det}^{k0}$ \label{sub:classification}}

\begin{figure}
\includegraphics[bb=0bp 0bp 595bp 842bp,height=1\textheight]{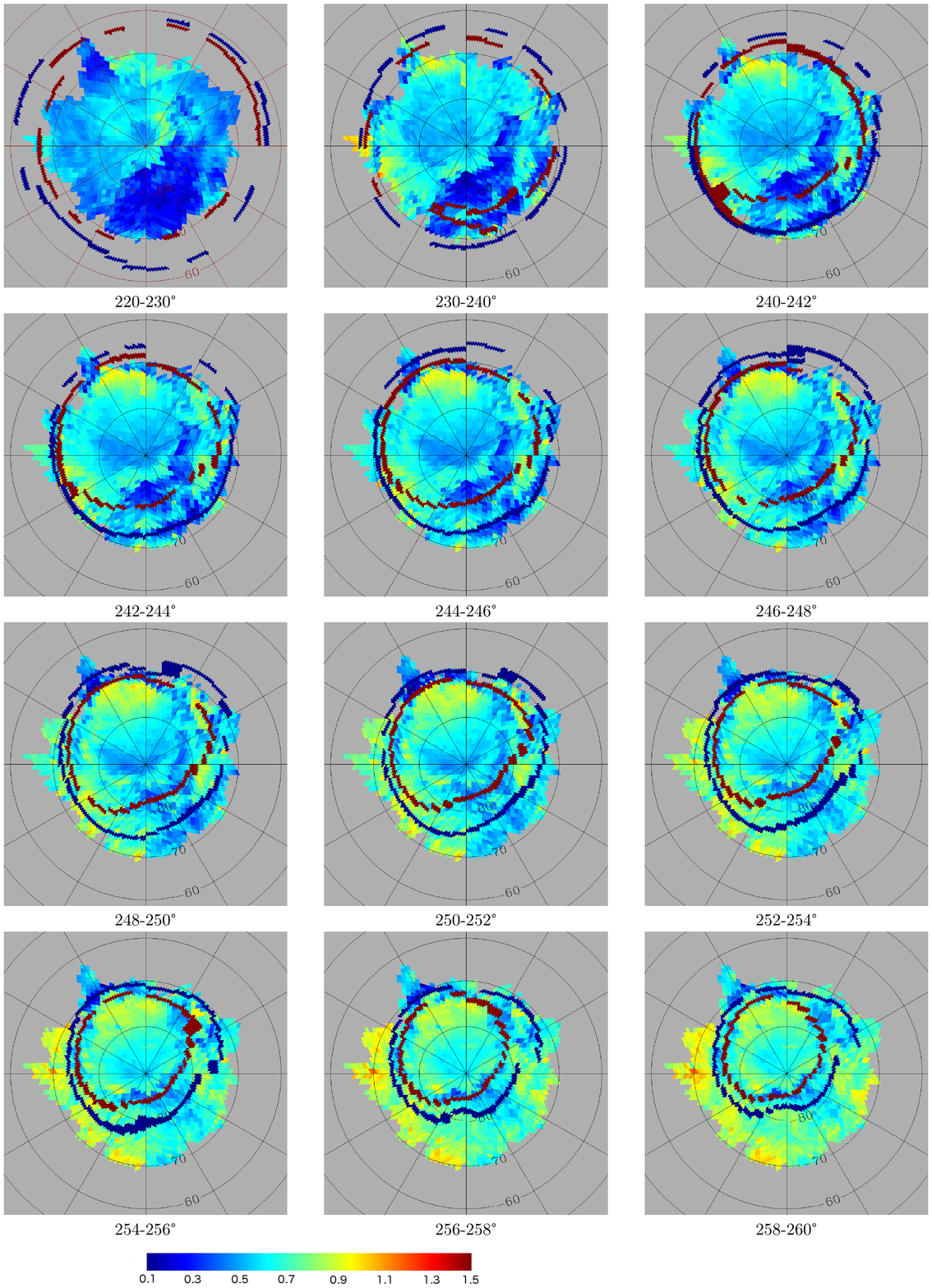}

\caption{Time series of orthographic mosaics depicting from \ls=220\textdegree{}
to \ls=260\textdegree{} the spatial distribution of seasonal mean
values for the aerosol optical depth at 1 \textmu{}m. The inner (respectively
outer) crocus line of the South Seasonal Polar Cap is coloured in
red (respectively in blue).\label{fig:mosaics_mod_1}}
\end{figure}

\begin{figure}
\includegraphics[bb=0bp 0bp 595bp 842bp,height=1\textheight]{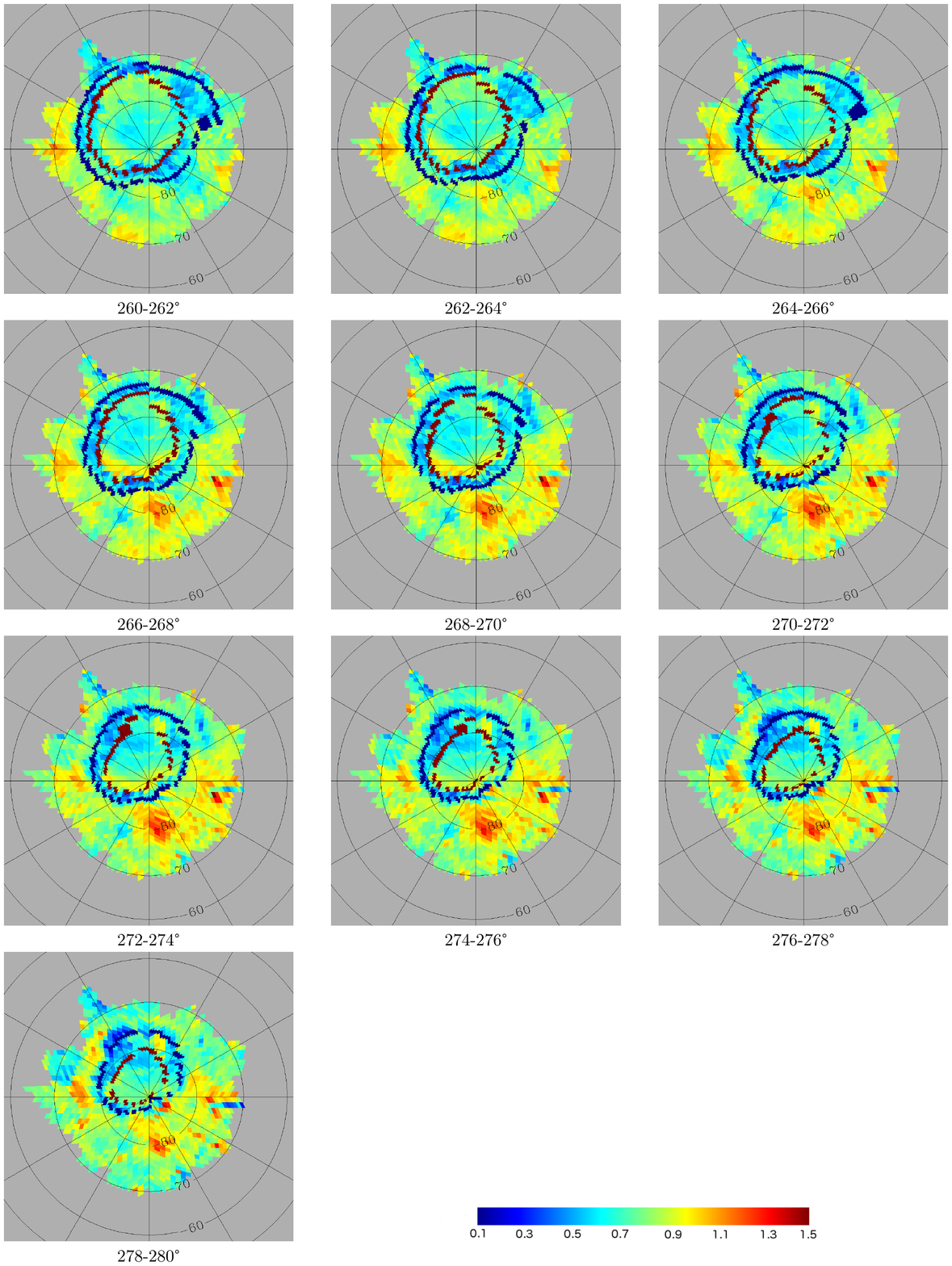}

\caption{Same as in Figure \ref{fig:mosaics_mod_1} but from \ls=260\textdegree{}
to \ls=280\textdegree{}. \label{fig:mosaics_mod_2}}
\end{figure}

\begin{figure}
\begin{tabular*}{1\textwidth}{@{\extracolsep{\fill}}c}
\includegraphics[bb=0bp 200bp 612bp 792bp,clip,width=1\textwidth]{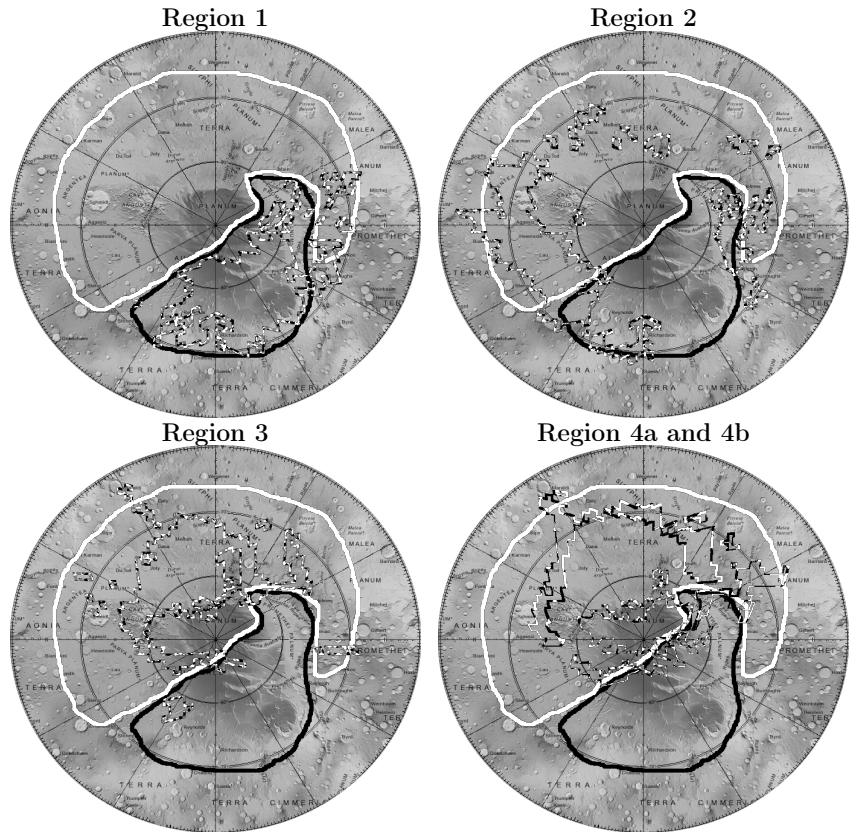}\tabularnewline
\tabularnewline
\end{tabular*}

\caption{Approximate spatial boundaries of the ``cryptic'' and ``anti-cryptic''
regions as well as of the four main classes of seasonal trends followed
by the AOD $\tau_{det}^{k0}$ in the southern polar region. The boundaries
appear respectively as plain black, plain white and striped contours.
For unit4 the width of the stripes is different for sub-units 4a (large)
and 4b (fine). See Sections \ref{sub:Main-spatio-temporal-units}
and \ref{sub:Regional-traits} for a detailed description.\label{fig:units_contours}}
\end{figure}

\begin{figure}
\includegraphics[bb=10bp 150bp 550bp 600bp,clip,height=0.3\paperheight]{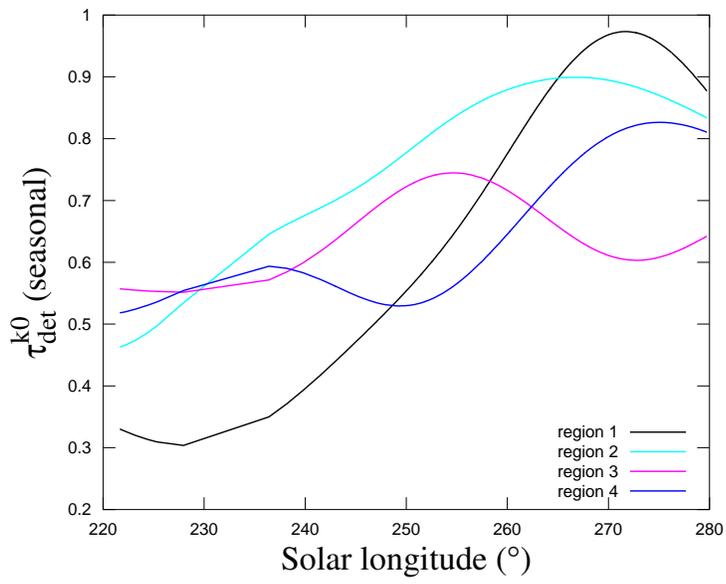}

\caption{Characteristic seasonal baseline of the four main classes of seasonal
trends followed by the AOD $\tau_{det}^{k0}$ in the southern polar
region. See Section \ref{sec:Trends} for a detailed description.\label{fig:units_baselines}}
\end{figure}

Even though the modelling of the seasonal trend is performed in a
pixel-wise manner, the spatial coherency of the time series of mosaics
is excellent as can be seen in Figures \ref{fig:mosaics_mod_1} and
\ref{fig:mosaics_mod_2}. Then we may expect that in the MIDA, baselines
can be gathered, based on similar shapes, into a limited number of
classes. To test this hypothesis the whole collection of baselines
is processed by kmeans classification provided by the R statistical
package (\url{http://www.r-project.org/}). The main difficulty to
overcome is the prior evaluation of the class number $N_{class}$.
For that purpose, several runs of the kmeans routine are conducted
independently with an increasing value for this unknown input parameter.
Then statistical tests \citet{Sugar2003} are performed through the
generation of a likelihood function depending on $N_{class}$ and
peaking at the most likely value $\hat{N}_{class}$ for the previous
parameter. We found $\hat{N}_{class}$=4 though with a poor separability
of the classes (inter-class variance only accounts for 32 \% of the
total variance of the data). This is to be expected when studying
atmospheric conditions that transition progressively at the global
scale of our investigation from one regime to the next. Nonetheless
Figures \ref{fig:units_contours} and \ref{fig:units_baselines} respectively
demonstrate that a good spatial coherency and seasonal baseline separability
is achieved for the four classes which boundaries must be considered
as indicative only.

\subsection{Assessing day-to-day variability and spatial heterogeneity \label{sub:mosaics_var}}

\begin{figure}
\includegraphics[height=0.9\textheight]{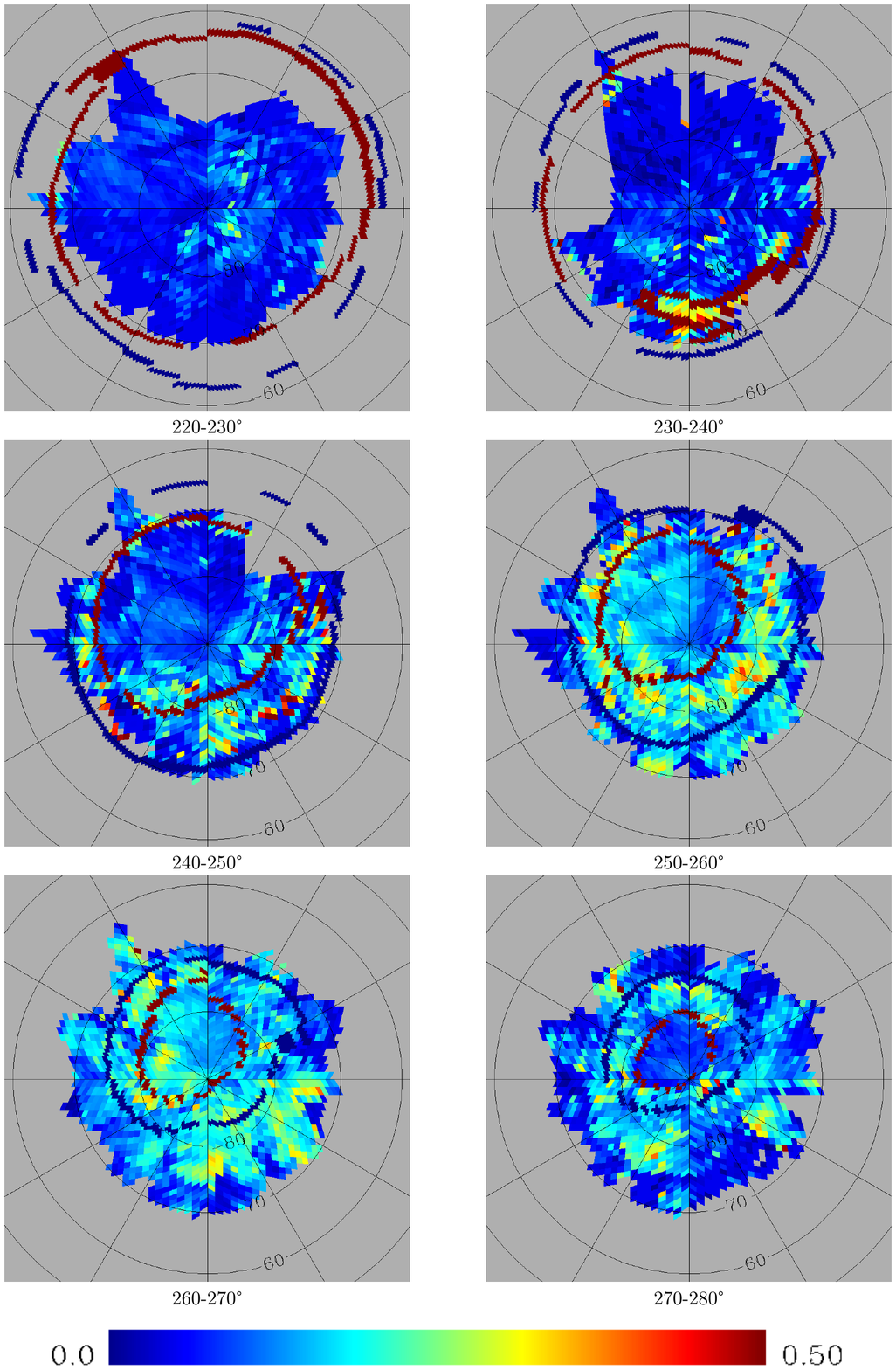}

\caption{Time series of orthographic mosaics depicting from $L_{s}$=220\textdegree{}
to $L_{s}$=280\textdegree{} the spatial distribution of day-to-day
variance values for the aerosol optical depth at 1 \textmu{}m. The
inner (respectively outer) crocus line of the SSPC is coloured in
red (respectively in blue).\label{fig:mosaics_var}}
\end{figure}

\begin{figure}
\begin{tabular*}{1\textwidth}{@{\extracolsep{\fill}}c}
\includegraphics[bb=10bp 150bp 550bp 600bp,clip,width=1\textwidth]{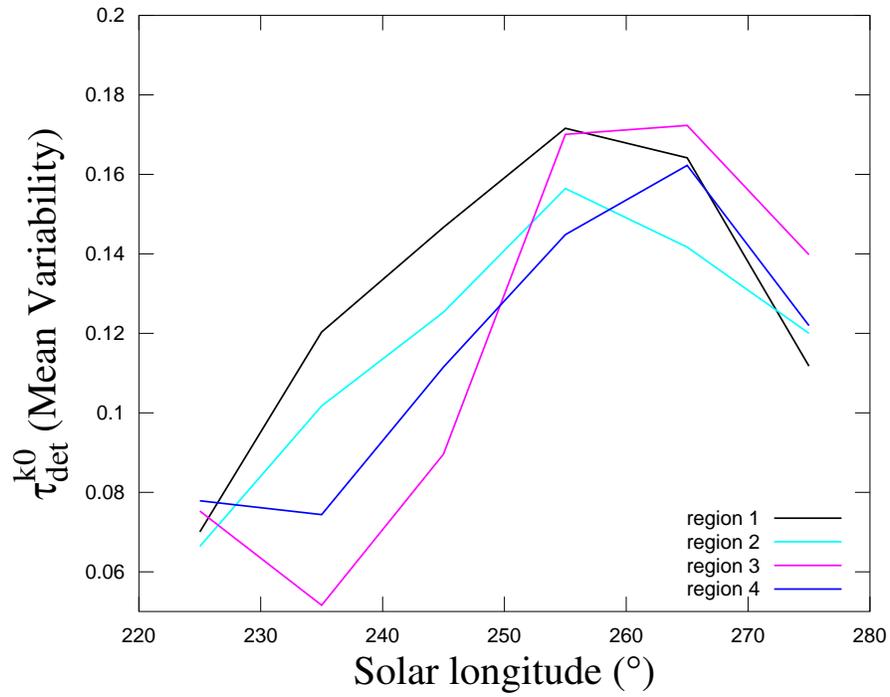}\tabularnewline
\includegraphics[bb=10bp 150bp 550bp 600bp,clip,width=1\textwidth]{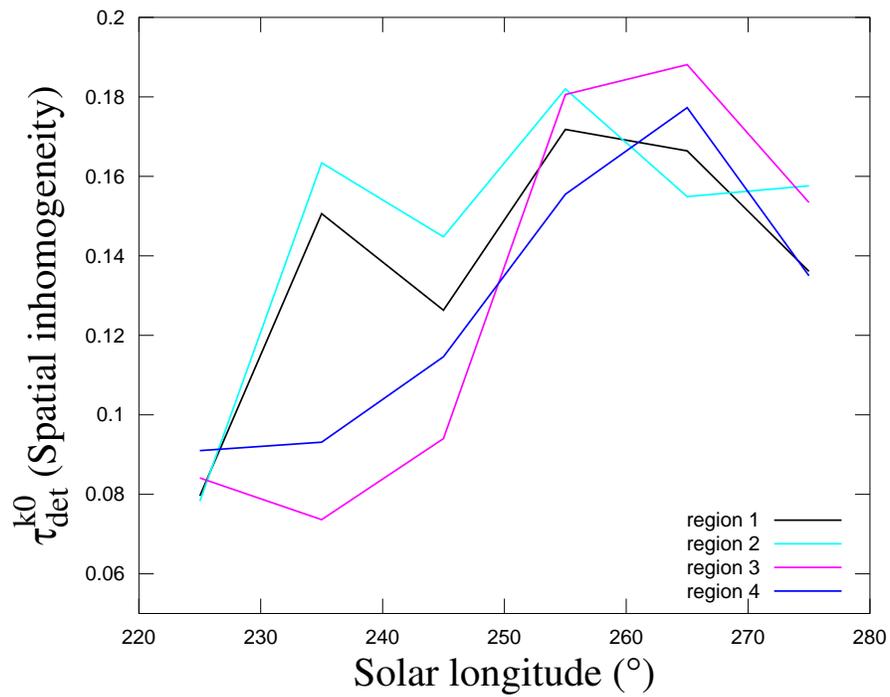}\tabularnewline
\end{tabular*}

\caption{For each of the four spatio-temporal units, the average seasonal evolution
of the day-to-day variability (up) and the seasonal evolution of spatial
heterogeneity (down).\label{fig:var_study}}
\end{figure}

In Section \ref{sub:describing-time-evolutions} we define the variable
component as the difference between the original signal and the mean
trend respectively extracted from the PIDA and MIDA for the same (x,y)
coordinates. We simply define an estimator of the day-to-day variability
magnitude by calculating for each valid HEALPix bin the root mean
square (RMS) of these differences over non-overlapping contiguous
intervals of solar longitudes 10\textdegree{} in duration. By mapping
the estimator in the geographical space according to the orthographic
projection, six mosaics are obtained that indicate the most and least
active locations for each time period (Figure \ref{fig:mosaics_var}).
The relevance of the mosaics has been checked by comparing them to
conjugated maps indicating the integrated number of original image
pixels that falls into each bin of the PIDA for \ls=220-230\textdegree{},
230-240\textdegree{}, 240-250\textdegree{}, 250-260\textdegree{},
260-270\textdegree{}, and 270-280\textdegree{}. No correlation is
found for any of the time periods. In addition to producing mosaics,
a spatial averaging of the estimator is performed separately over
the four spatial regions conjugated to the classes distinguished in
the previous Section resulting in distinct temporal trends (Figure
\ref{fig:var_study} upper graph). As regards to spatial heterogeneity,
we perform its assessment over the same four spatial regions. For
that purpose we first calculate the root mean square of the variable
component over a given region at each date, - time index - of the
PIDA. We then obtain four temporal curves at full time resolution
that are smoothed by a sliding average operation (width of the window:
10\textdegree{} \ls). The result is then sampled at the solar longitudes
225\textdegree{}, 235\textdegree{}, 245\textdegree{}, 255\textdegree{},
265\textdegree{}, and 275\textdegree{} (Figure \ref{fig:var_study}
lower graph).

\section{Trends of atmospheric dust opacity for the high southern latitudes\label{sec:Trends}}

\subsection{Main spatio-temporal units\label{sub:Main-spatio-temporal-units}}

Figure \ref{fig:units_contours} displays the contours of the four
principal spatio-temporal units that have emerged from classifying
the collection of seasonal trends followed by the dust optical depth.
The first unit roughly spans the area between meridians 90\textdegree{}E
and 210\textdegree{}E and parallels 70\textdegree{}S and 85\textdegree{}S.
It corresponds basically to the portion of the southern polar layered
deposits (SPLD) known as the ``cryptic region''  (see definition
in Section \ref{sub:definitions-SSPC}). It is characterised by a
steady rise of $\tau_{det}^{k0}$ from $\approx$ 0.3 at the beginning
of the covered period to a maximum $\approx$ 1.0 at \ls=270\textdegree{}.
The rate of increase is moderate for $230\lesssim L_{s}\lesssim240\text{\textdegree\ }$,
before being accentuated. Once passed the maximum, one can observe
a noticeable drop in the curve. Unit 2 is composed basically of two
non contiguous area, the most extended being between meridians 210\textdegree{}E
and 300\textdegree{}E and parallels 70\textdegree{} and 80\textdegree{}:
Parva Planum, Argentera Planum. The eastern most part of the area
is actually slightly deported towards the lower latitudes. The least
extended area of unit 2 is centred around (78\textdegree{}S,100\textdegree{}E;
western part of Promethei Planum). As regards to its temporal behaviour,
this unit can be distinguished by a steady rise of $\tau_{det}^{k0}$
that starts at higher values than unit 1 ($\approx$ 0.5), has a significantly
lower rate of increase, and reaches a maximum value of $\approx$
0.9 ( at \ls=265\textdegree{}). Then $\tau_{det}^{k0}$ declines
to some degree. Unit 3 is quite contiguous even though it displays
a complex shape. The major part of it lies principally between longitudes
270\textdegree{}E and 30\textdegree{}E and at latitudes as high as
87\textdegree{}S and as low as 70\textdegree{}S including Dorsa Argentera.
It also comprises the area occupied by the so-called Mountain of Mitchell
located near 70\textdegree{}S, 40\textdegree{}E \citep{James2000}
and an ``arm'' of SPLD around the south permanent cap at longitudes
between 120\textdegree{}E and 270\textdegree{}E. Units 2 and 3 basically
coincide with the anti-cryptic region. The atmospheric opacity of
unit 3 is already quite high at the beginning of our period of interest
around $\approx$ 0.55, stays quite constant until \ls=240\textdegree{},
and increases afterwards up to a maximum of $\approx$ 0.75 at \ls=255\textdegree{}.
Then it decreases down to a minimum of $\approx$ 0.6 at \ls=272\textdegree{}.
Our curve indicates the beginning of a reversal afterwards, but it
is not possible to say if it is significant. Finally unit 4 surrounds
the quasi totality of unit 3 as a ``stripe'' passing through (clockwise)
the permanent cap and the nearby SPLD, the Argentera Planum, the Sisyphi
Planum, and the Dorsa Brevia. A striking anti-correlation between
the temporal behaviours of unit 3 and 4 can be noticed. Until \ls=240\textdegree{},
the dust opacity remains basically at the same level ($\approx$0.55)
despite moderate rise for unit 4. Afterwards the baseline of the latter
reaches its minimum ($\approx$ 0.5 at \ls=250\textdegree{} ) when
the baseline of unit 3 reaches its maximum ($\approx$ 0.75) and the
other way around ($\approx$ 0.8 at \ls=275\textdegree{} for unit
4, $\approx$ 0.6 for unit 3).

\subsection{Regional traits of AOD activity\label{sub:Regional-traits}}

The spatio temporal units with their distinct temporal signatures
are the summarised expression of the dust loading of the atmosphere
 depicted with more details by the time series of mosaics built from
the MIDA (explained in Section \ref{sub:mosaics} and displayed in
Figures \ref{fig:mosaics_mod_1} and \ref{fig:mosaics_mod_2}). In
addition one can appreciate the  regression of the seasonal deposits
by looking at the simultaneous displacement of the inner and outer
crocus lines.

\paragraph*{Cryptic region}

When the latter region, which is nearly in a one to one correspondence
to unit 1, is entirely covered by \co\ frost ($220\lesssim L_{s}\lesssim230\text{\textdegree\ }$),
the overlying atmosphere AOD is the lowest of the entire seasonal
cap. As the inner crocus line progresses through the region toward
the high latitudes - faster than in any other longitude sectors -
patches of more dusty atmosphere appear above the area that are just
defrosted. By \ls=254\textdegree{}, only the central part of the
cryptic region displays AOD values below $\approx$ 0.5. Superimposed
on the mean behaviour of steadily increasing opacity as noted above,
we observe by \ls=262\textdegree{} the emergence of a regional AOD
maximum centred at (80\textdegree{}S,165\textdegree{}E) that grows
in value and spatial extent and reaches its height at \ls=272-274\textdegree{},
and starts to disappear afterwards but not completely. Thus this regional
enhancement of dust opacity is present for at least $\approx$ 20
days. In the most western part of the cryptic region around the location
(78\textdegree{}S,100\textdegree{}E; western part of Promethei Planum),
there is an area of the same sort but that has been active even earlier
(\ls=230\textdegree{}). As result it is classified as belonging to
unit 2 rather than to unit 1.

\paragraph*{Anti-cryptic region}

Unit 2 which falls principally in the latter region sees in many places
a rise of atmospheric opacity since the beginning, i.e. even before
the passage of the crocus lines. Defrosting has only a modest influence
on the rise and several regional maxima's develop. The most important
one is along longitude 270\textdegree{}E and migrates by $\approx$5\textdegree{}
towards the higher latitudes  between \ls=252\textdegree{} and \ls=280\textdegree{}.
Unit 3 is the portion of the seasonal cap that also shows relatively
high and growing values of the overlying atmosphere AOD at least for
$240\lesssim L_{s}\lesssim255\text{\textdegree\ }$. However contrary
to unit 2, this trend then starts to reverse with the crossing of
the inner crocus line. Consequently the subsequent prevalence of ice-free
terrains seems correlated with a substantial apparent clearing of
the atmosphere. At the end of the period of interest we observe a
regional minimum of dust opacity around (75\textdegree{}S,330\textdegree{}E;
Dorsa Argentera). We distinguish two sub-units for unit 4 (see Figure
\ref{fig:units_contours}). At the lowest latitudes that we cover
in the ``anti-cryptic'' sector (unit 4a) an early increase of AOD
(modest in amplitude) followed by a decline is also observed linked
to the passage of the inner crocus line. After \ls=250-270\textdegree{}
with the passage of the outer crocus line, the AOD starts to recover
reaching $\approx$ 0.5 to 1 at the end of the period. Curiously,
even though the region of the permanent cap (unit 4b) does not experience
the crossing of the crocus lines, it displays a very comparable AOD
seasonal baseline and thus  has been also classified in unit 4 by
the kmeans algorithm. Nevertheless this might just be a coincidence.
To conclude one striking feature of the anti-cryptic region is thus
the nearly continuous strip of relatively low AOD values ($\approx$0.5-0.6)
that accompanies the progressing transition zone between the inner
and outer crocus lines after \ls=262-264\textdegree{}. Such a propagating
front explains the major characteristics of the temporal behaviour
of units 3 and 4 and their anti-correlation.

\subsection{Studying day-to-day variability and spatial heterogeneity\label{sub:Studying-day-to-day-variability}}

In Section \ref{sub:mosaics_var} an estimator of the magnitude of
the day-to-day variability is put forward and mapped at a resolution
of 1.0\textdegree{}.pixel$^{\text{-1}}$ for six non-overlapping contiguous
intervals of solar longitudes 10\textdegree{} in duration (220-230\textdegree{},
230-240\textdegree{}, 240-250\textdegree{}, 250-260\textdegree{},
260-270\textdegree{}, and 270-280\textdegree{}; Figure~\ref{fig:mosaics_var}).
The area where the resulting mosaics display meaningful values is
more restricted in general than the area for which we have models
of the mean trend of $\tau_{det}^{k0}$ versus time. Indeed calculation
of the estimator requires even more OMEGA observations per time intervals
than for the model. Nevertheless this is not a limiting factor for
studying the day-to-day variability poleward of 70\textdegree{}S latitude.
For $220\lesssim L_{s}\lesssim230\text{\textdegree\ }$ the variability
is low ($\approx$0.07) throughout the SSPC with a few exceptions.
Then, passed \ls\ $\approx$ 230\textdegree{}, the variability begins
to rise very significantly (up to 0.3) in the ``cryptic'' sector
at large (longitudes 60-240E\textdegree{}) regardless of the relative
position of the considered bin with the inner crocus line. During
the next time interval, such signs of intense low scale dust activity
have intensified in the previous sector but have also propagated both
clockwise and anti-clockwise. In the anti-cryptic sector enhanced
values for the variance are confined within the transition zone. Inside
the SSPC only the 0-60\textdegree{}E interval of longitudes remains
relatively quiet. Then we reach the period \foreignlanguage{french}{(}$250\lesssim L_{s}\lesssim270\text{\textdegree\ }$)
during which high variability is widespread among all sectors. Finally
\foreignlanguage{french}{(}$270\lesssim L_{s}\lesssim280\text{\textdegree\ }$)
the dust activity decreases in intensity especially inside the remaining
SSPC and for latitudes equatorward of $\approx$ 75\textdegree{}S.
Figure \ref{fig:var_study} illustrates the average seasonal evolution
of the day-to-day variability (upper panel) and the seasonal evolution
of spatial heterogeneity (lower panel) for each of the four units
distinguished in Section \ref{sub:classification}. These graphs corroborate
the previous examination. Furthermore they reveal that for $220\lesssim L_{s}\lesssim250\text{\textdegree\ }$
each spatio-temporal unit has a very distinct signature. The atmospheric
dust cover above the ``cryptic'' region (unit 1) displays the highest
temporal variability in conjunction of being the second most spatially
segregated. Nevertheless a look at Figure \ref{fig:units_baselines}
 reminds us that the absolute level of AOD is in general relatively
lower than anywhere else. At the contrary above the ``anti-cryptic''
sector including the ``Mountains of Mitchell'' (unit 3) the dust
cover is much more spatially uniform also implying a low day-to-day
variability. According to the latter indicator alone, units 2 and
4 are intermediate. After \ls\ $\approx$ 255\textdegree{} all units
behave comparably to a first approximation.

\subsection{Tracking local time dependencies\label{sub:Tracking-local-time}}

\begin{figure}
\includegraphics[bb=0bp 0bp 595bp 842bp,height=0.9\textheight]{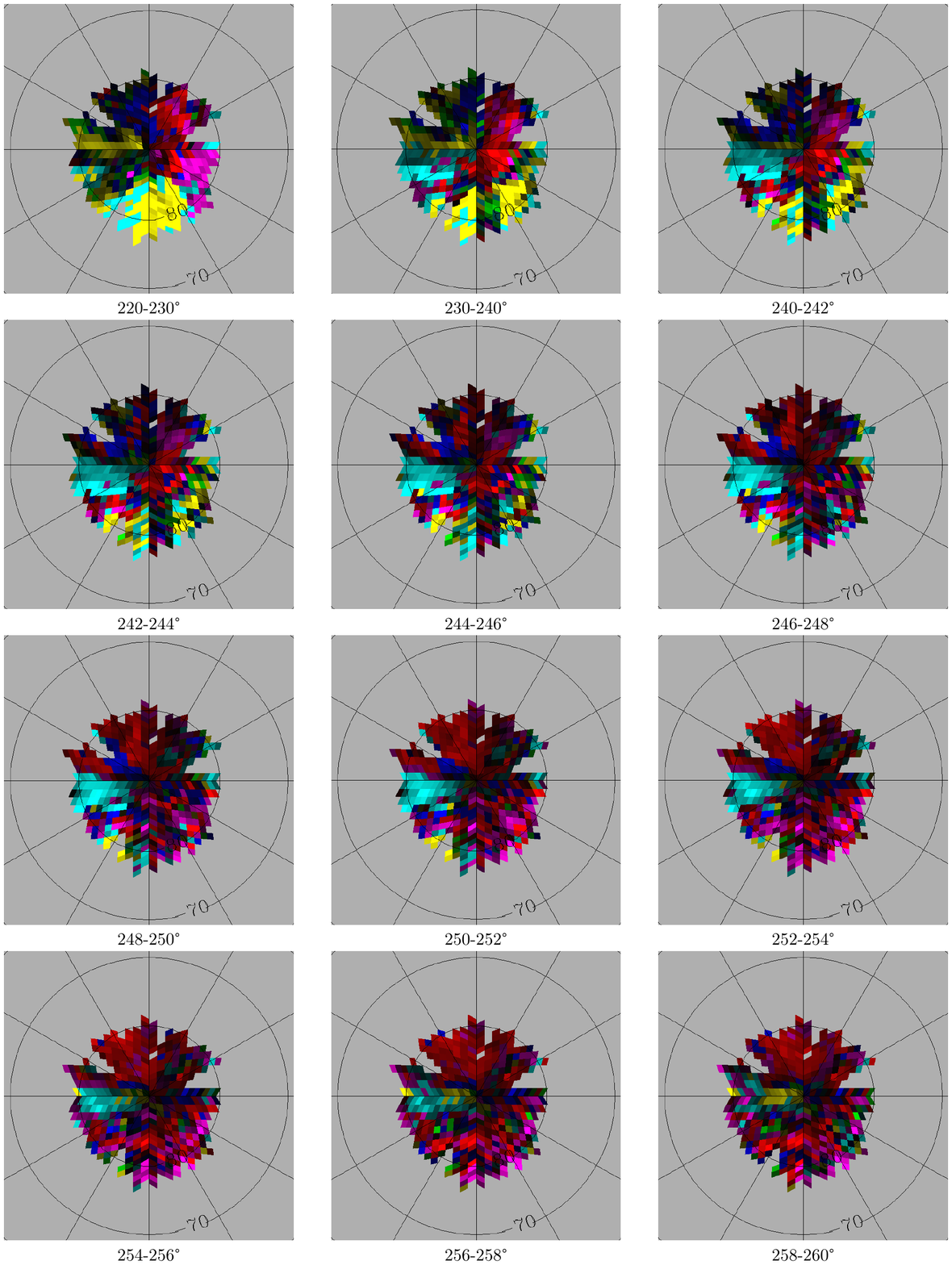}

\caption{Time series of orthographic mosaics depicting from \ls=220\textdegree{}
to \ls=260\textdegree{} the spatial distribution of local time dependency
for the aerosol optical depth at 1 \textmu{}m. We assign to each bin
a distinctive primary hue and a luminosity depending on the ordering
of the triplet of LT $\tau_{det}^{k0}$ values and on its variance
$var_{LT}$ as follows : \foreignlanguage{english}{1: }red $B_{0-6}<B_{6-12}<B_{12-18}$,
2: magenta $B_{6-12}<B_{0-6}<B_{12-18}$, 3: blue $B_{0-6}<B_{12-18}<B_{6-12}$,
4: cyan $B_{6-12}<B_{12-18}<B_{0-6}$, 5: green $B_{12-18}<B_{0-6}<B_{6-12}$,
6: yellow $B_{12-18}<B_{6-12}<B_{0-6}$. Colour bar appears in Fig.
\ref{fig:mosaics_lt2}. \label{fig:mosaics_lt1}}
\end{figure}

\begin{figure}
\includegraphics[bb=0bp 0bp 595bp 842bp,height=0.9\textheight]{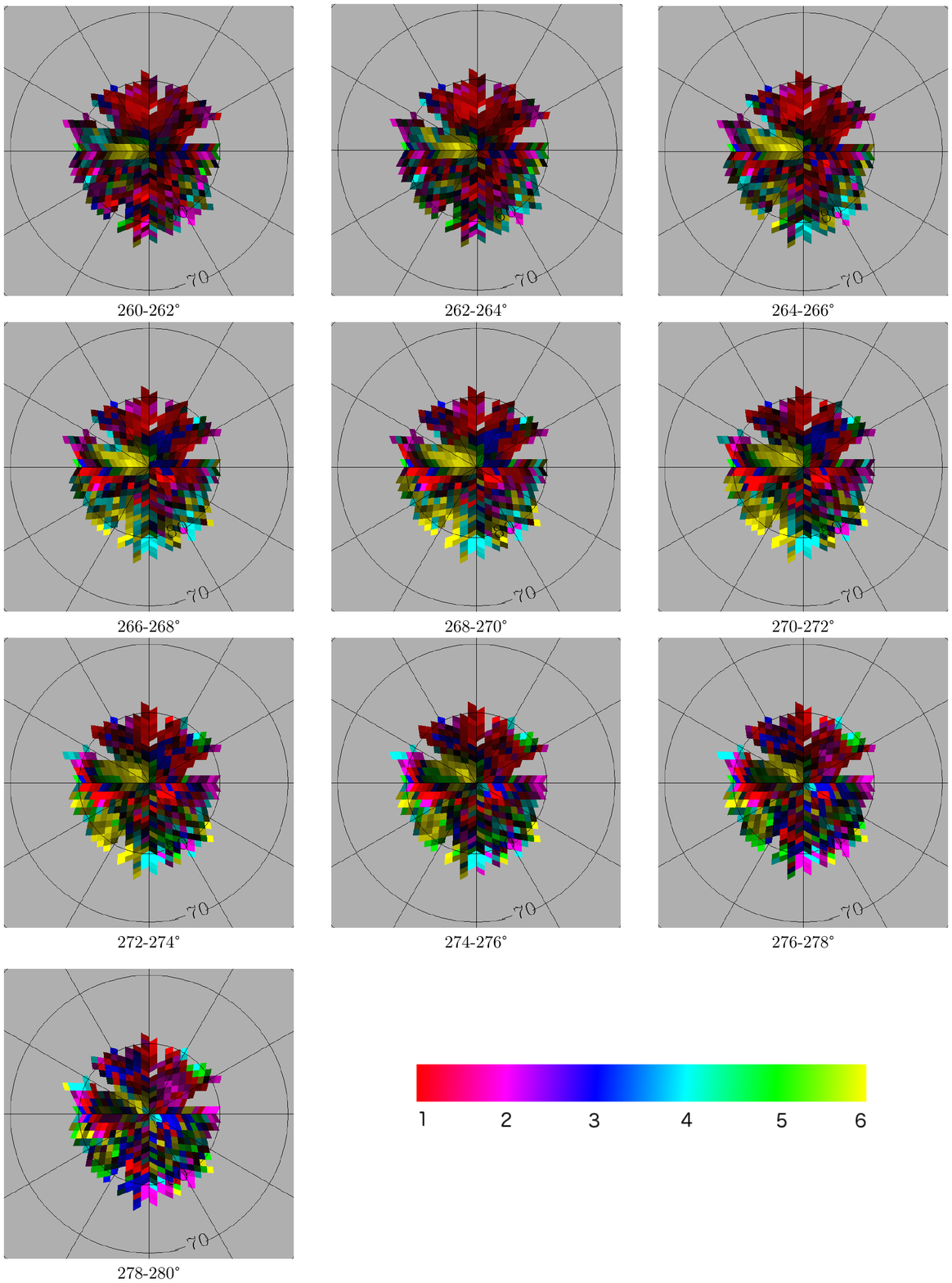}

\caption{Same as in Figure \ref{fig:mosaics_lt1} but from \ls=260\textdegree{}
to \ls=280\textdegree{}. \label{fig:mosaics_lt2}}
\end{figure}

In this section our goal is to characterise local time dependencies
of atmospheric optical depth during spring and beginning of summer
for the high southern latitudes. In Section \ref{sub:map_Integration}
we mention that an additional dimension can be optionally added to
the PIDA by considering the local time of acquisition for each OMEGA
pixel. A division of the martian sol into three equal Local Time (LT)
intervals is adopted: 0-6, 6-12, and 12-18. Identically to what is
described in Section \ref{sub:describing-time-evolutions} we model
the mean trend of $\tau_{det}^{k0}$ versus \ls\ although each LT
interval is now treated independently. As a result any systematic
shift affecting the AOD correlated to the diurnal cycle should be
dissociated from the variability associated to random $\tau_{det}^{k0}$
estimation errors and day-to-day meteorology for example. The drawback
of adding the LT dimension is that the number of samples per bin of
the HEALPix grid is now much reduced. The bins for which enough points
are available and sufficiently distributed in the period of interest
fall predominantly poleward of the 80$^{\text{th}}$ parallel, the
area that was scrutinised the most frequently by OMEGA. Finally the
modelling leads to the generation of three MIDA, one per LT interval.
Prior to further processing, we average along the Y dimension of each
array all the valid values falling in the reference solar longitude
intervals: \ls=220-230\textdegree{}, 230-240\textdegree{}, 240-242\textdegree{},
etc. At this point quantifying the spread and the ordering of the
triplet of seasonal $\tau_{det}^{k0}$ values $(B_{0-6},B_{6-12},B_{12-18})$
- one per baseline - is performed for each eligible bin and each \ls\
range. In addition a better visualisation of the results is achieved
by means of a special colour coding. We assign to each bin a distinctive
primary hue and a luminosity depending on the ordering of the triplet
and on its variance $var_{LT}$ as follows. Red, magenta, blue, cyan,
green, and yellow correspond respectively to the six possible ordering,
i.e. permutations, among the three baselines. The luminosity is a
continuous real number in the {[}0,1{]} interval that is set to $\min(\sqrt{var_{LT}}/0.22,1)$.
The denominator of the previous fraction represents three times the
standard deviation of $\sqrt{var_{LT}}$ when doing the statistics
on the whole dataset. By mapping our colour coded joint estimators
of ordering and variance in the geographical space according to the
orthographic projection we obtain a time series of mosaics (Figures
\ref{fig:mosaics_lt1} and \ref{fig:mosaics_lt2}). Bright and subdued
colours indicate significant, respectively undetectable, LT variability.
Table \ref{tab:colors-AOD-LT} summarises the local time pattern of
dust activity for each color regime effectively realised in the series
of mosaics.

\begin{table}
\begin{tabular}{|c|c|}
\hline 
Colour regime & Local time pattern of dust activity\tabularnewline
\hline 
\hline 
Yellow & $B_{12-18}<B_{6-12}<B_{0-6}$\tabularnewline
\hline 
Magenta & $ $$B_{6-12}<B_{0-6}<B_{12-18}$\tabularnewline
\hline 
Red & $B_{0-6}<B_{6-12}<B_{12-18}$\tabularnewline
\hline 
Cyan & $B_{6-12}<B_{12-18}<B_{0-6}$ \tabularnewline
\hline 
\end{tabular}

\caption{Correspondence between the main colours appearing in Figures \ref{fig:mosaics_lt1}
and \ref{fig:mosaics_lt2} and the local time pattern of dust activity.\label{tab:colors-AOD-LT}}
\end{table}

At the beginning of the covered period \ls=220-230\textdegree{} a
clear dichotomy separates the ``cryptic sector'' of longitudes with
high diurnal variance and the ``anti-cryptic'' sector with much
less pronounced values of $var_{LT}$. Nevertheless in both cases
we observe two regimes: $ $$B_{6-12}<B_{0-6}<B_{12-18}$ (magenta)
in the western part (0-150\textdegree{}E) and $B_{12-18}<B_{6-12}<B_{0-6}$
(yellow) in the eastern part (150-300\textdegree{}E). The remaining
area shows no sign of diurnal variability until \ls $\approx$ 240\textdegree{}.
During the period $230\lesssim L_{s}\lesssim244{^\circ}$ the portion
of unit 1 (see Section \ref{sub:classification}) covered by the map
stays predominantly in the yellow regime even though the middle part
shows very low values of $var_{LT}$. Meanwhile closer to the pole
over a quite extended area we rather have $B_{0-6}<B_{6-12}<B_{12-18}$
(``red'' regime). In the interval $244\lesssim L_{s}\lesssim248{^\circ}$
a transition can be observed. Then for $248\lesssim L_{s}\lesssim262{^\circ}$
the predominance of the red regime is obvious everywhere except in
the sector 240-270\textdegree{}E where the ordering remains in the
regime $B_{6-12}<B_{12-18}<B_{0-6}$ (cyan) until $L_{s}\lesssim258{^\circ}$.
In any case the variance is moderate to low. After \ls $\approx$
264\textdegree{} and at least until \ls $\approx$ 274\textdegree{}
the yellow regime returns with quite high diurnal variability over
the SPPC and for longitudes 180-240\textdegree{}E at the outskirts
of the covered area. Past this date no clear spatially coherent trends
are visible in the mosaics in terms of ordering nor in terms of spread.

\subsection{Individual features\label{sub:Transient-events}}

\begin{figure}
\includegraphics[bb=45bp 80bp 700bp 600bp,clip,width=1.4\textwidth]{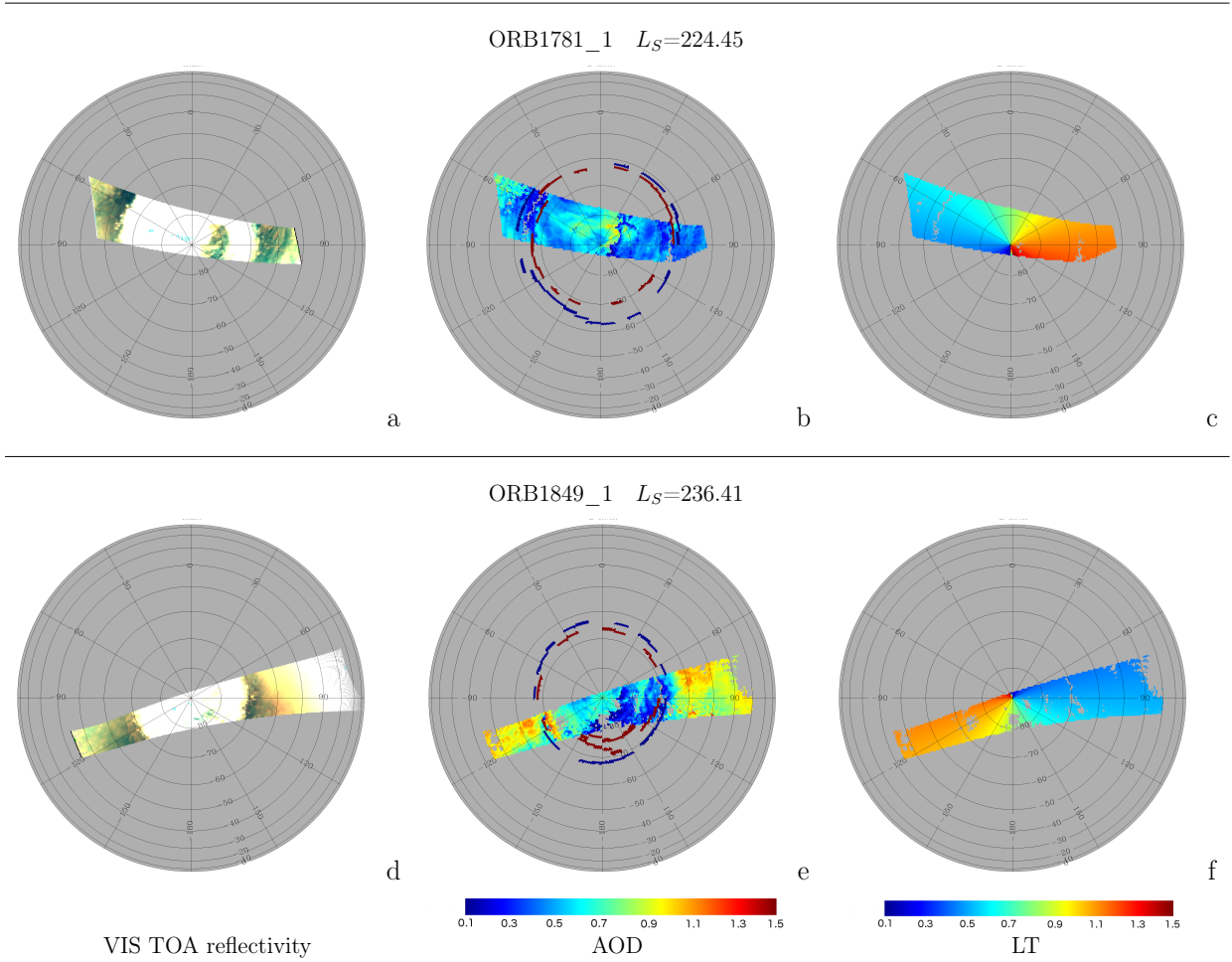}

\caption{A selection of two global OMEGA observations and associated products.
The left column displays an RGB composition of the TOA martian reflectivity
in the visible which is stretched so as to reveal dust in the atmosphere
as yellowish hues. The central column displays the Aerosol Optical
Depth map at 1 micron. The right column displays a map which colour
scale is used to indicate the local time of pixel acquisition. See
the text for details.\label{fig:event0}}
\end{figure}

\begin{figure}
\includegraphics[bb=60bp 20bp 842bp 580bp,clip,width=0.8\paperwidth]{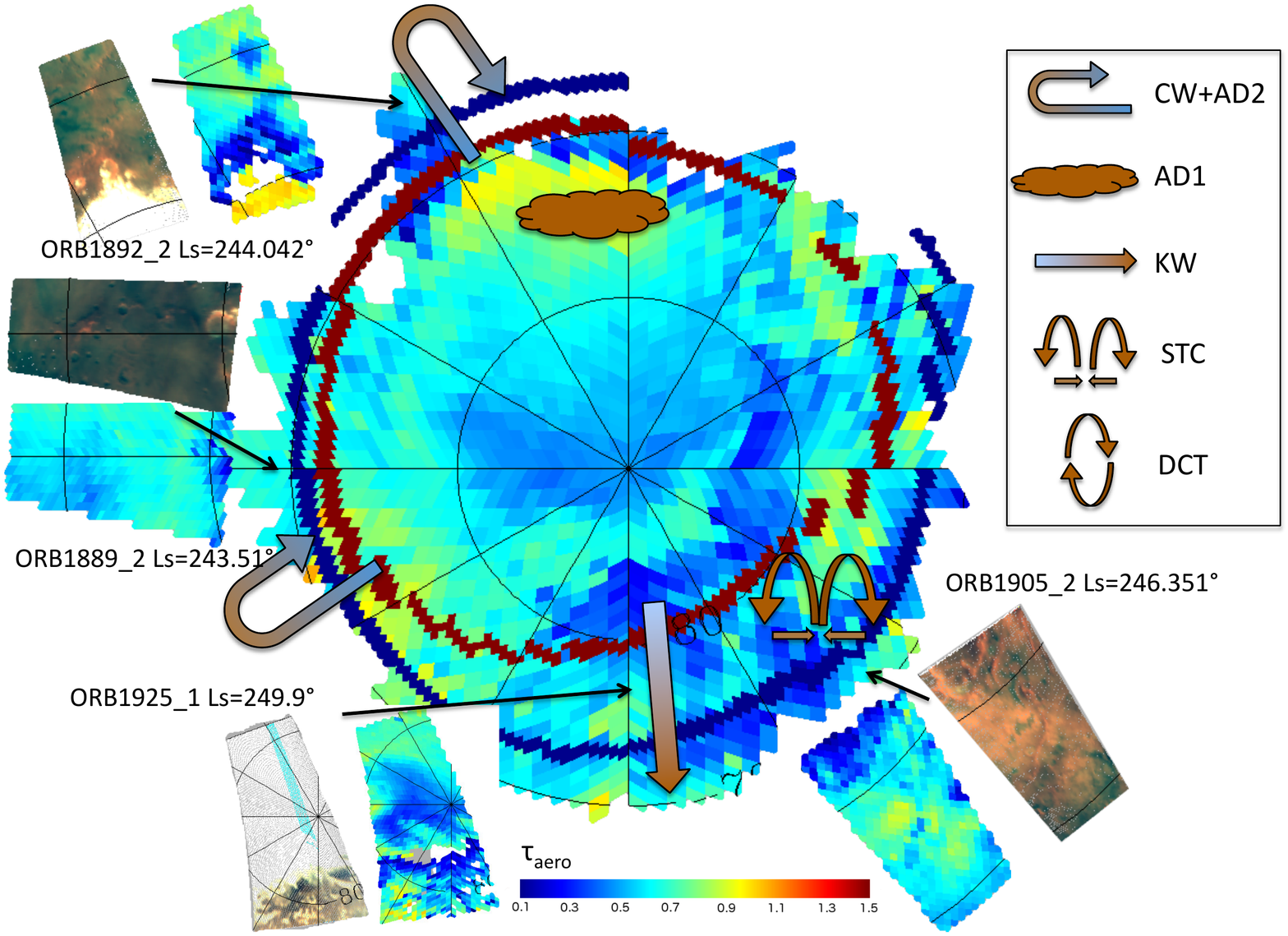}

\caption{Composition aimed at illustrating and supporting the discussion. The
global map of the seasonal level of AOD for Ls=244-246\textdegree{}
serves as the background and also gives the position of the two crocus
lines. Markers are superposed that indicate the possible occurrence
of the different dynamical mechanisms mentioned in the text for the
period Ls=240-250\textdegree{}. In the legend, AD1 and AD2 stand for
advection of dust respectively by high altitude return flows and cap
winds (CW). KW are katabatic winds, DCT daytime convective turbulence
in the boundary layer, and STC small scale thermal circulation in
the transition zone. Finally a selection of pairs of products derived
from individual OMEGA observations are presented and put in their
geographical context. The RGB composition is the top-of-atmosphere
martian reflectivity at three visible channels which is stretched
so as to reveal dust in the atmosphere as yellowish hues. The map
is the corresponding Aerosol Optical Depth at 1 micron. See the text
for details. \label{fig:event1}}
\end{figure}

\begin{figure}
\includegraphics[bb=60bp 20bp 842bp 580bp,clip,width=0.8\paperwidth]{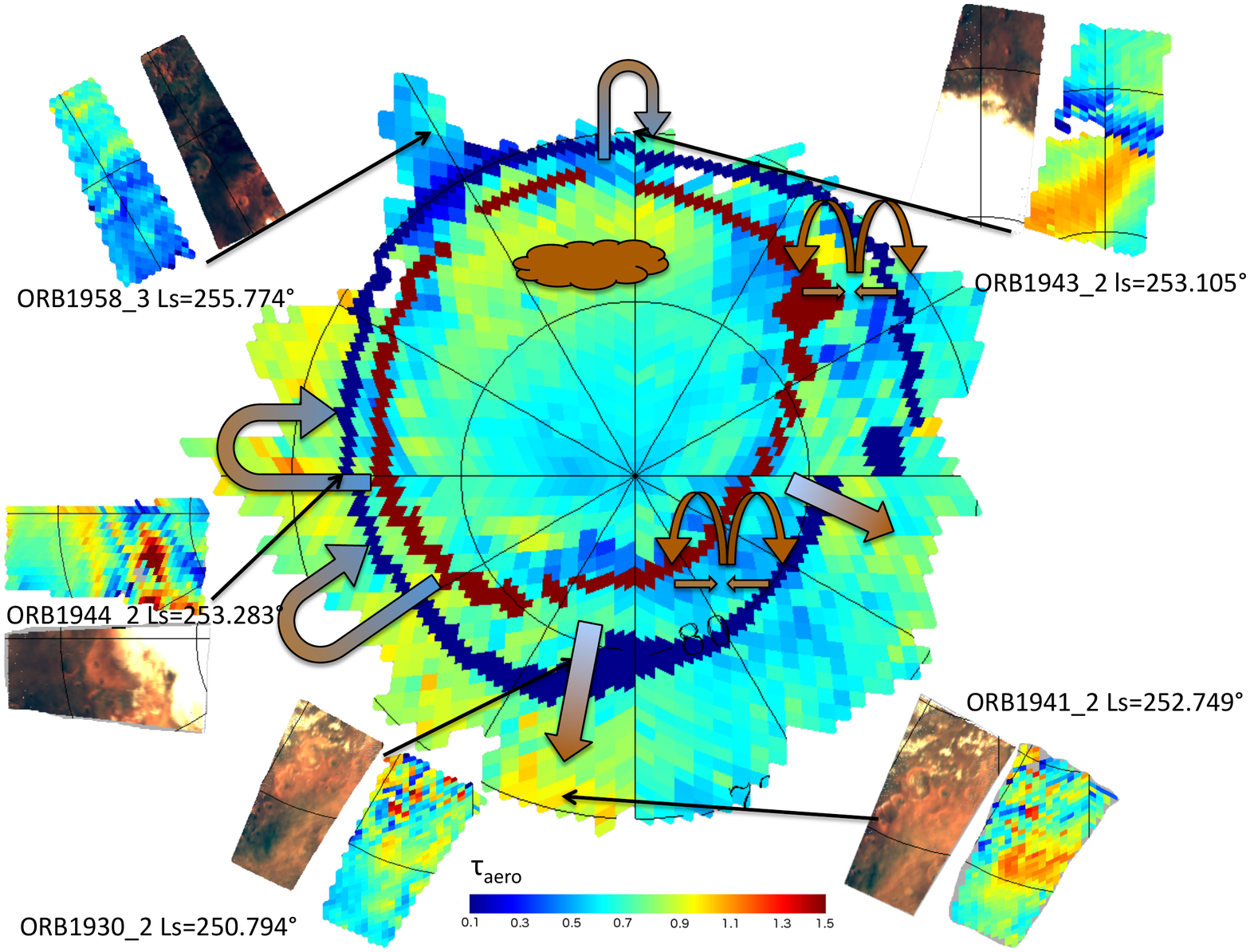}

\caption{Same as Fig. \ref{fig:event1} but for the period Ls=250-260\textdegree{}\label{fig:event2}}
\end{figure}

\begin{figure}
\includegraphics[bb=60bp 20bp 842bp 580bp,clip,width=0.8\paperwidth]{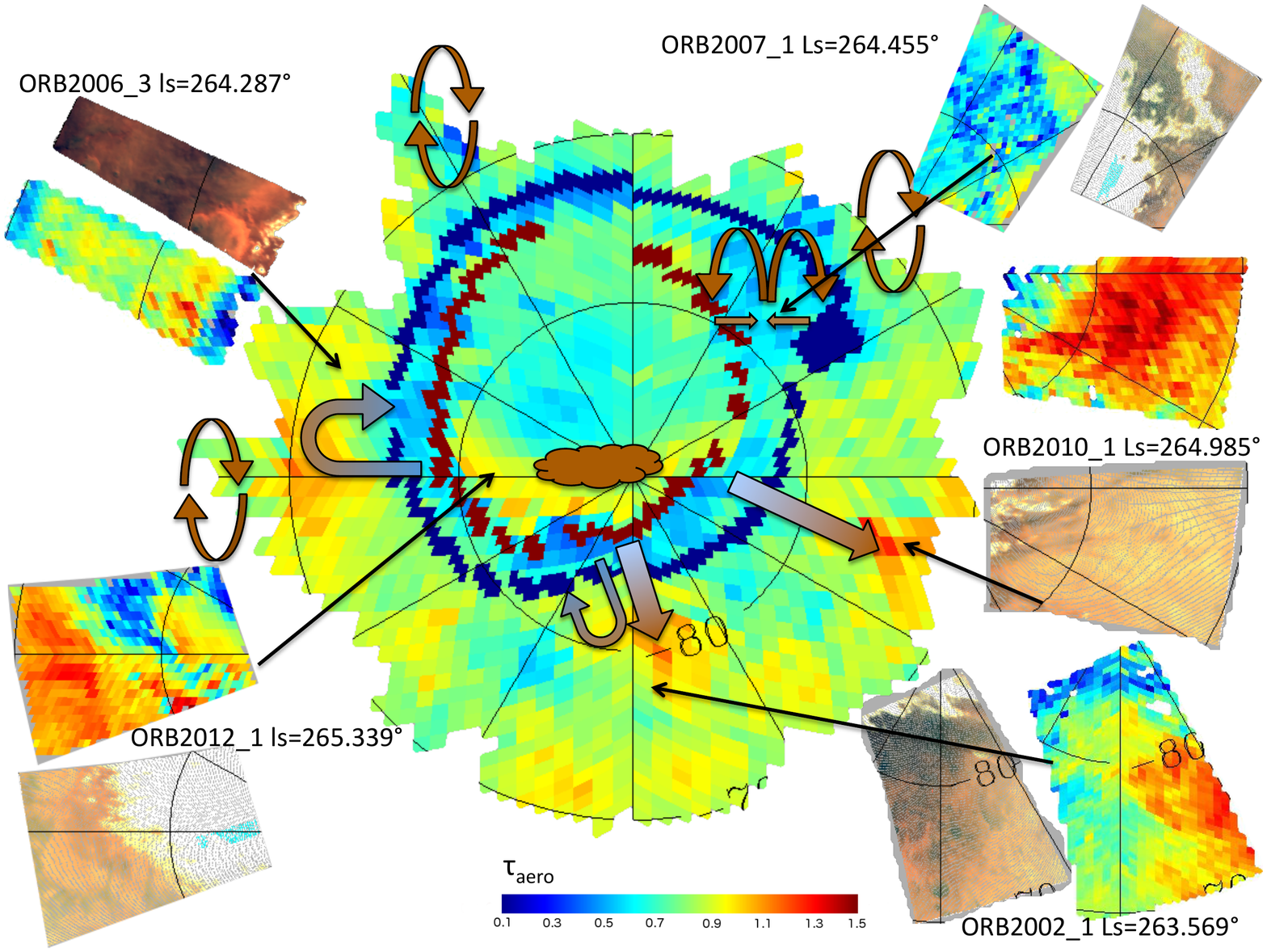}

\caption{Same as Fig. \ref{fig:event1} but for the period Ls=260-270\textdegree{}.\label{fig:event3}}
\end{figure}

\begin{figure}
\includegraphics[bb=60bp 20bp 842bp 580bp,clip,width=0.8\paperwidth]{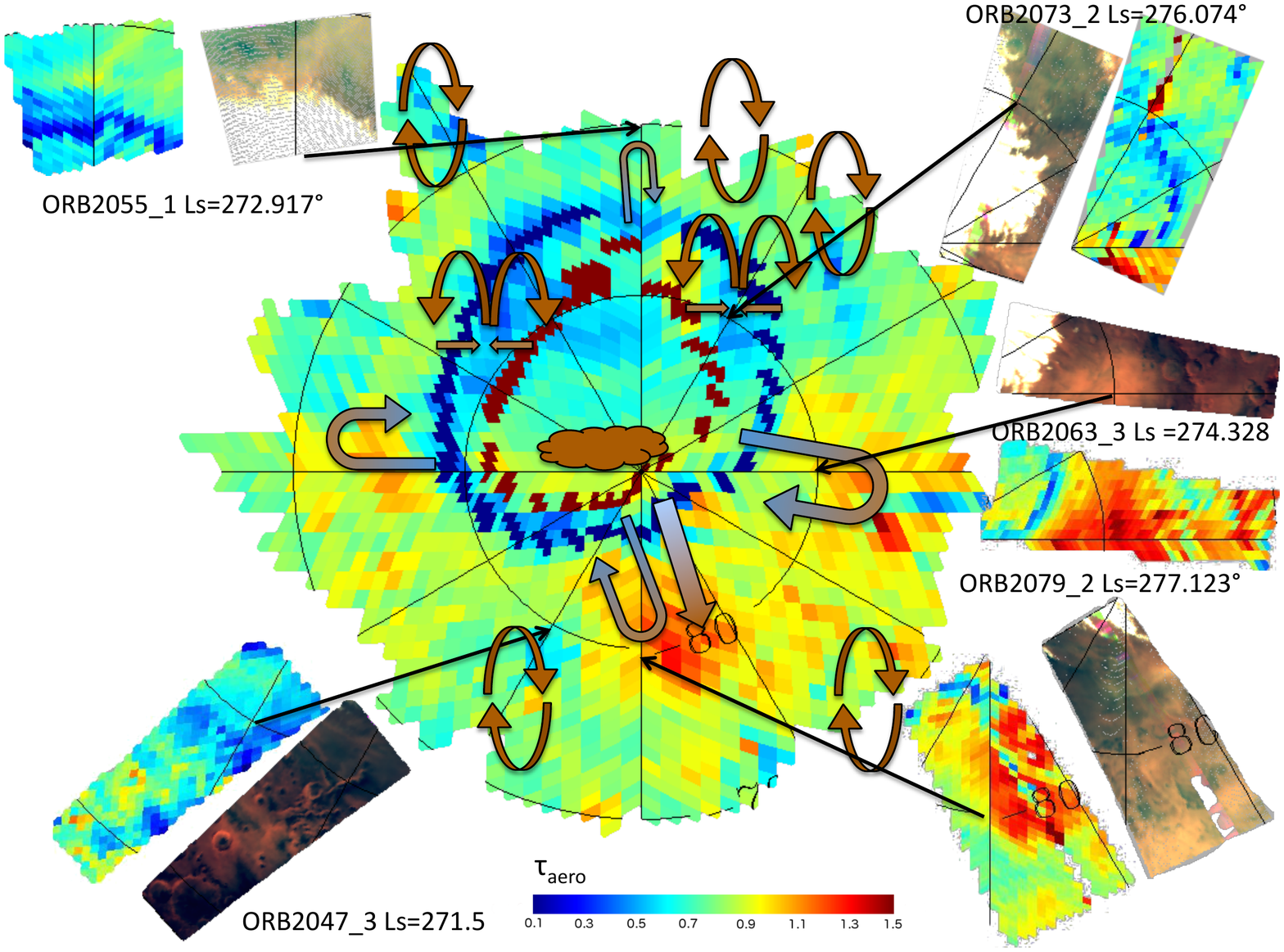}

\caption{Same as Fig. \ref{fig:event1} but for the period Ls=270-280\textdegree{}.\label{fig:event4}}

\end{figure}

Further insights into the dust activity going on around the cap edge
and at lower latitudes emerges from the examination of individual
projected AOD maps (Section \ref{sub:mosaics}). Information provided
by each map is enhanced by providing two companion maps associated
to the same observation. In the first, an RGB composition of the top-of-atmosphere
(TOA) martian reflectivity measured by OMEGA at three wavelengths
(0.7070, 0.5508, and 0.4760 \micron) is shown. It is stretched so
as to make visible the dust clouds as yellowish hues against mineral
surfaces. Nevertheless it should be noted that such a stretch completely
saturates the image over the the seasonal deposits. In the second
map, a colour scale is used to indicate the local time of pixel acquisition.
Figures \ref{fig:event0} to \ref{fig:event4} show a selection of
nineteen observations out of the initial collection of 284 observations
that illustrate different types of representative situations. In Figure
\ref{fig:event0} each observation corresponds to a line in a Table,
the RGB composition, the AOD map, and the LT map being respectively
arranged in the first, second and third column. In the other four
Figures, each observation corresponds to the RGB composition and the
AOD map alone, both superimposed on a context mosaic depicting the
level of seasonal AOD. The pair is identified as well as situated
in time (solar longitude Ls) and space (arrows). The following description
is based on the entire collection of individual observations but illustrated
with the selection. 

In the interval $L_{S}\approx$ 220-242\textdegree{} for which OMEGA
provided global scale observations, we first note significant dust
opacity's around the cap ($\tau_{det}^{k0}\approx$0.6-0.7) on the
``morning'' side and modest values on the ``afternoon'' side of
the AOD maps (Fig.\ref{fig:event0}b,c). This is corroborated by the
visible images (Fig.\ref{fig:event0}a). Then starting around $L_{S}\approx$
230\textdegree{} and lasting a least until $L_{S}\approx$ 240\textdegree{}
both the AOD maps (Fig.\ref{fig:event0}e) and their visible counterparts
(Fig.\ref{fig:event0}d) reveal thick dust clouds mostly in the vicinity
of the seasonal cap regardless of the local time (Fig.\ref{fig:event0}f).
It is worth mentioning that while the clouds are well outside the
cap on the ``morning'' side, they penetrate well passed the transition
zone on the ``afternoon'' side. After $L_{S}\approx$ 240\textdegree{}
until the end of the covered period,  observations turn to more frequent
but more spatially focused snapshots. The interval $L_{S}\approx$
242-255\textdegree{} is a period of usually quiescent activity in
the afternoon for the defrosted area of units 1, 2, and 4a with low
to moderate level of opacity ($\tau_{det}^{k0}\approx$ 0.2-0.5; ORB1889\_2,
Fig.\ref{fig:event1}, ORB1943\_2, ORB1958\_3 Fig.\ref{fig:event2}).
In the visible such a situation means a clear view of the surface
brown features and low TOA reflectivities.  Sometimes events of higher
intensity ($\tau_{det}^{k0}$ up to $\approx$ 1.0) can be spotted
(ORB1892\_2, ORB1905\_2, Fig.\ref{fig:event1}). In the large transition
zone of unit 1 the dust opacity retrieved by our method remains low
while significant opacity is apparent in the visible images (ORB1925\_1,
Fig.\ref{fig:event1}). A similar situation occurs at $L_{S}\approx$
250-275\textdegree{} for longitudes 0 to 60\textdegree{}E, unit 3
(ORB2007\_1, Fig.\ref{fig:event3}) From at least $L_{S}\approx$
250\textdegree{} to the end of the covered period ($L_{S}\approx$
280\textdegree{}) marked opacity can be recognised around the cap
with clear patterns depending on the spatio-temporal unit that this
is considered. Once passed the crocus line, unit 1 (``cryptic region'')
presents area of high atmospheric opacities that persist regardless
of the local time. Nevertheless the highest values of $\tau_{det}^{k0}$
occur more frequently during the second part of the ``night'' and
in the morning (ORB2002\_1, ORB2010\_1, ORB2012\_1, Fig.\ref{fig:event3},
and ORB2079\_2, Fig.\ref{fig:event4}). Then surface features in the
visible are subdued behind a diffuse veil of scattering aerosols with
relatively elevated TOA reflectivities. Note that sometimes the development
of the dust cloud is readily appreciable both in the AOD map and the
RGB composition as in the pair of observations ORB1930\_2, ORB1941\_2,
Fig.\ref{fig:event2} both acquired during the afternoon.  For regions
2 and 4a the area close to defrosting area undergo locally intense
dust activity ($\tau_{det}^{k0}\approx$1-1.5) that are observed rather
during the afternoon hours than during the early morning (ORB1944\_2,
Fig.\ref{fig:event2}, ORB2006\_3, Fig. \ref{fig:event3}, ORB2055\_1,
Fig. \ref{fig:event4}). A typical transient event captured by ORB2063\_3
- a ``storm'' of local proportions - is shown in Fig. \ref{fig:event4}.
One last observation driven by the examination of the dataset concerning
units 2 and 4a is that the area that had been defrosted for $\approx$10\textdegree{}
of solar longitude show low to moderate AOD corresponding to quite
clear conditions in the visible images (ORB2047\_3, Fig. \ref{fig:event4}).

\subsection{Comparison with earlier works\label{sub:comparison}}

In this last sub-section, we confront briefly the evolution and variability
of the AOD given by on the one hand our study (OMEGA images in MY
27) and, on the other hand, by the study of \citet{Toigo2002,Imamura2011}
(TES and MOC images in MY 24, 25 and 26 ).

Basically \citet{Toigo2002} present two (TES) maps of dust opacity
at 9 \textmu{}m. In addition they also discuss a discontinuous series
of MOC daily global mosaics in color. The first TES map depicts the
situation  at Ls $\approx$ 228\textdegree{} in which infrared opacity
can be in excess of 0.5 all around the edge of the polar cap while
it remains between 0.1 and 0.2 at lower latitudes (background level).
This is the expression in the infrared of what is depicted more qualitatively
but with more spatial details by a MOC mosaic. Typically a dust cloud
is created near the edge of the polar cap, is blown away from the
cap while growing in extent, and then dissipates over a period of
6-8 hours. In particular at \ls$\approx$228\textdegree{} a number
of discrete well-defined dust clouds can be seen everywhere around
the cap edge, especially in the 180-270\textdegree{}E longitude sector.
Our AOD maps generated from individual observations taken by OMEGA
at the beginning of the period of interest ($220\lesssim L_{s}\lesssim242\text{\textdegree) }$
cover large strips of Mars, joining across the cap areas of opposite
longitudes at moderate latitudes (Section \ref{sub:Transient-events}).
The regions external to the cap (latitudes equatorward of $\approx$
70\textdegree{}S that are not visible in our mean trend mosaics, Fig.\ref{fig:event0})
indeed display extended cloud systems following \ls $\approx$ 225\textdegree{}
for varied longitude sectors, regardless of local time. Nonetheless
due to the limited time and spatial coverage of the subsequent OMEGA
observations, it is not possible to monitor the propagation of transient
dust clouds from the cap edge to the lower latitudes.

In the second TES map at Ls $\approx$ 257.5\textdegree{} significant
accumulation of dust is restricted to the longitudes from 270 to 45\textdegree{}E.
At contrary the most productive area in our case around \ls $\approx$
255-260\textdegree{} of MY 27 is in the longitude sector 180-300\textdegree{}E
(unit 2), rather in agreement with the results of \citet{Imamura2011}
(see below). 

The work of \citet{Imamura2011} led to a more systematic mapping
of the dust at 9 \textmu{}m. For Martian Year (MY) 24, Ls=270-280\textdegree{},
a global map displays an elongated dusty region that encircles the
south pole with a gap in the range 330-60\textdegree{}E; this feature
was also observed in MY 25 and 26. The map is part of a larger study
that also addresses the temporal and spatial variation of the 9 \textmu{}m
dust optical depth in this region using Hovmöller diagrams for three
Martian years. For the range of solar longitudes $240\lesssim L_{s}\lesssim300$,
\citet{Imamura2011} found that dust clouds emerge repeatedly (every
10-20 sols) from an area delimited by latitudes 70–80\textdegree{}S
and longitudes 240–300\textdegree{}E. Then they move westward and
reach the region in the longitude sector 60–120\textdegree{}E to finally
disappear. The overall longitude range of the disturbance, 60–300\textdegree{}E,
coincides with elevated terrains in the south polar region, and with
the increase of dust optical depth observed at Ls=270-280\textdegree{}
that encircles the south pole. The quasi-periodic behaviour (period
of 10–20 sols) of the dust optical depth disturbance that is recognised
in the latitude band 70-80\textdegree{}S by \citet{Imamura2011} in
their Hovmöller diagrams is absent from our curves of seasonal AOD
even for the main source regions (Unit 2 until \ls$\approx$250\textdegree{},
then Unit 1). 

To summarise, even though similar broad tendencies arise from the
different studies at certain dates and for certain regions, discrepancies
are the rule. Inter-annual variability and notable differences in
terms of resolutions can account for this situation.

\section{Discussion\label{sec:Discussion}}

\begin{figure}
\includegraphics[bb=55bp 0bp 790bp 850bp,clip,height=0.8\textheight]{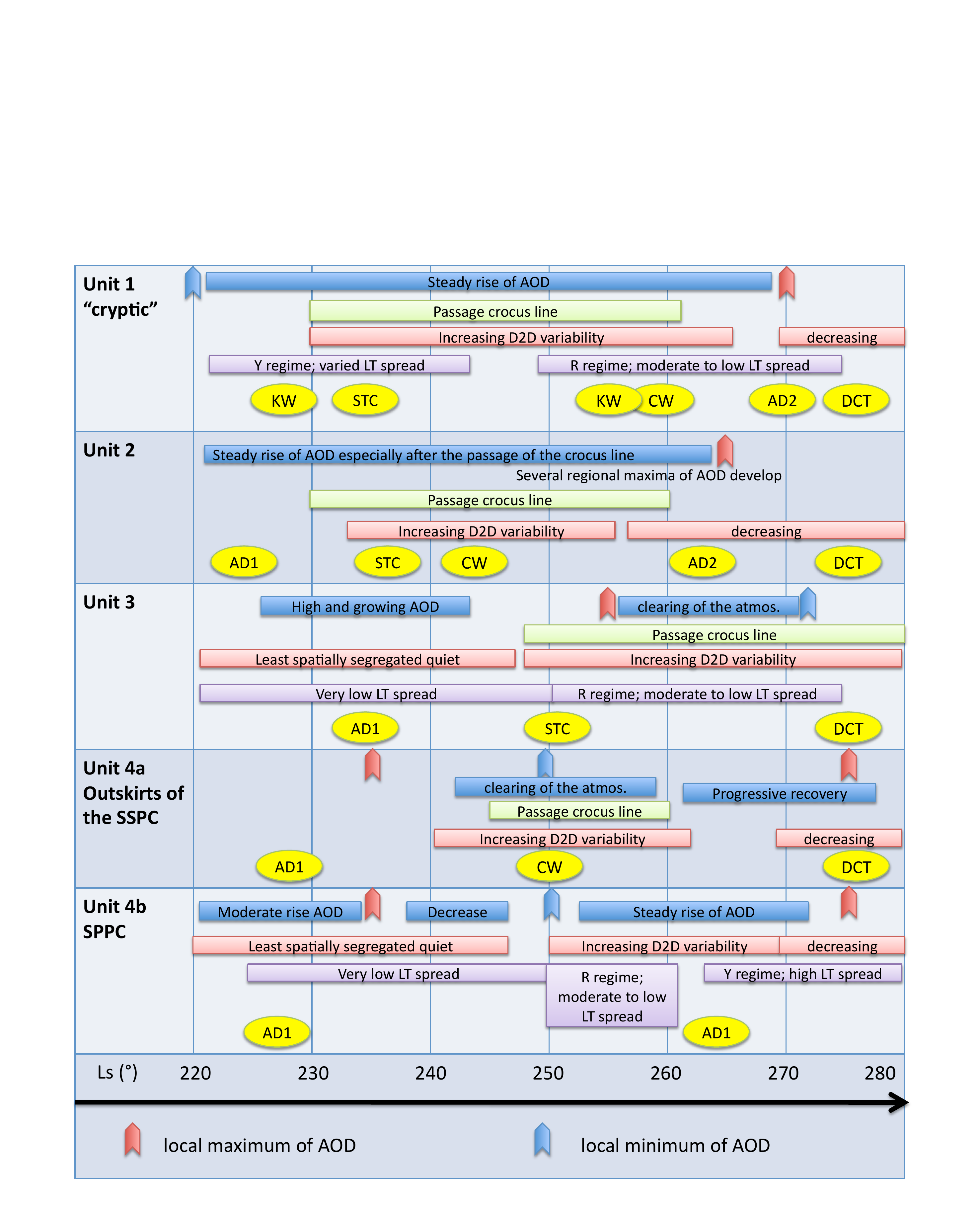}

\caption{Synthetic view of the atmospheric dust activity in the high southern
latitudes of Mars in mid-spring to summer for MY27. AOD Aerosol Optical
Depth. AD1 and AD2: advection of dust respectively by high altitude
return flows and cap winds (CW). KW: katabatic winds. DCT: daytime
convective turbulence in the boundary layer. STC small scale thermal
circulation in the transition zone. Y and R respectively yellow and
red regimes. D2D: day to day. The markers of local maxima/minima of
the AOD are related to those of Figure~\ref{fig:units_baselines}
.\label{fig:Synoptic-view}}
\end{figure}

Figure \ref{fig:Synoptic-view} proposes a qualitative summary of
the principal results obtained in Section \ref{sec:Trends}. The four
main spatio-temporal units are followed in parallel according to the
solar longitude. In addition we indicate which time interval is concerned
by the passage of the crocus lines. A series of indicators - the seasonal
trend of $\tau_{det}^{k0}$ (Section \ref{sub:describing-time-evolutions}),
the day-to-day variability $var_{D2D}$ and spatial heterogeneity
(Section \ref{sub:mosaics_var}), as well as the ordering and spread
$var_{LT}$ of the triplet of baseline values $(B_{0-6},B_{6-12},B_{12-18})$
(Section \ref{sub:Tracking-local-time}) are reported. In the present
section these indicators are now conjointly discussed.  The goal here
is to propose scenarios, i.e. succession in time of mechanisms or
conditions that plausibly control the atmospheric dust activity for
different area of the high southern latitudes in spring. First we
introduce several mesoscale and microscale circulation processes based
on earlier works. Then the expression, expected conditions of occurrence,
and dust lifting capabilities of such processes are phenomenologically
linked with the evolution and variability of the AOD observed with
OMEGA in conjunction with the state of the regressing seasonal cap.
This is the method by which we put forward hypotheses regarding which
circulation process(es) likely control the atmospheric dust activity
for a given unit and for a given time range. These proposed scenarios,
which are illustrated by a series of figures (\ref{fig:event1} to
\ref{fig:event4}), could be tested by future high-resolution atmospheric
mesoscale modelling.

\subsection{Introducing several mesoscale and microscale circulation processes.
\label{sub:Introducing-circulations}}
\begin{description}
\item [{Katabatic~Winds}] (KW) are drainage atmospheric flows that form
in the lowest atmospheric levels ($\approx$ one hundred meters) when
cooled dense air is accelerated down sloping terrains by gravity,
overcoming the opposing along-slope pressure gradient. Martian conditions
conducive to this behaviour are the extremely low temperatures of
the seasonal deposits in thermodynamical equilibrium with the overlying
\co\ gas and nighttime radiative cooling of surface and atmosphere.
In both cases they are maximum at night hours when the temperature
inversion above surfaces is the highest \citep{Spiga2011,Kauhanen2008}. 
\item [{Thermal~Circulations}] are triggered by lateral temperature heterogeneities
due to soil thermophysical contrasts but most importantly due to icy
vs. bare surface contrasts. They can cause high surface friction and
vertical mixing thus injecting dust in the atmosphere \citep{Spiga2010,Siili1999}.
Atmospheric flow modellings \citep{Holton2004} demonstrate that a
rough proportionality exists between the typical size of the temperature
heterogeneities and the height reached by the convection cells. Note
that the latter own a dynamical surface branch oriented from the lowest
to the highest temperatures and a return flow, i.e. blowing in opposite
direction, at higher altitude. In particular \citet{Toigo2002} have
studied the development of the so-called cap winds (CW), analogues
to terrestrial sea breeze circulation, generated by the strong thermal
contrasts that exist along the retreating edge of the southern spring
polar cap during the middle to late spring. Their three-dimensional
numerical modelling at mesoscales demonstrate that surface wind stresses
are sufficient at some location during specific local time intervals
to initiate movement of sand-sized particles and hence dust lifting
over quite large area. Thus they recognise cap edge winds, a mesoscale
structure ($\approx10-100$ km), as an important factor for the development
of dust storms near the cap edge. However the influence of a having
a gradual transition between the inner and outer crocus lines instead
of a neat cap edge and the availability of dust and/or sand particles
are not addressed in their paper. 
\end{description}
Note that \citet{Siili1999} performed a set of numerical experiments
with the two-dimensional Mars Mesoscale Circulation Model that shed
some new light about the previous two processes. In particular they
demonstrate that strong near surface winds potentially capable of
lifting dust from the surface can be achieved (i) in daytime along
a slope which lower section is frozen and upper section is ice-free:
``anabatic winds'' (ii) in nighttime over a fully ice-covered slope
or defrosted in its lower section: ``katabatic'' winds. Inclusion
of atmospheric dust ($\tau$ = 0.3) reduces the daytime ice-edge forcing
- the upslope flow is attenuated - while the nocturnal downslope flow
is enhanced. Consequently thermal circulations, in the form of anabatic
winds, can be enhanced by local relief in the defrosting area. Interestingly,
spatial segregation of frozen and ice free terrains in this transition
zone can be controlled by topography through slope orientation and
profile for example. 
\begin{description}
\item [{Daytime~Convective~Turbulence}] (DCT) is particularly well developed
in the Martian boundary layer during the afternoon under the influence
of the heated surface. Even in situations of weak large-scale and
mesoscale winds, this thermally-driven convection might itself cause
significant vertical mixing and wind gustiness. Two worth-mentioning
manifestations of convective turbulence are “convective gusts” found
near walls of convective cells and convective vortices linked with
``dust devils''.
\item [{Synoptic~Scale~Circulation}] at large-scale (>100s kilometers)
is characterised by inter-hemispheric meridional Hadley Cells and
mid-latitude baroclinic waves, the latter features occasionally accompanied
by dust fronts extending over thousands of kilometers.
\end{description}

\subsection{Unit 1}

For unit 1, that well corresponds to the ``cryptic'' region, we
distinguish four successive phases of activity. 

In the very early phase of defrosting ($220\lesssim L_{s}\lesssim240\text{\textdegree\ }$),
the seasonal AOD is the lowest of the entire seasonal cap, and we
have $B_{12-18}<B_{6-12}<B_{0-6}$ (i.e. a ``yellow regime'') with
dispersed values of $var_{LT}$, meaning that, statistically, there
is more dust in the atmosphere during the early hours than during
the morning and the afternoon with local variations of the spread.
For $220\lesssim L_{s}\lesssim230\text{\textdegree\ }$ the mantle
of seasonal \co\  deposits is still quite continuous and exhibits
a very compact texture in the form of a translucent slab \citep{Kieffer2000}.
Nonetheless it is then superficially contaminated by a large amount
of dust making it barely detectable by its spectral signature in the
shortwave infrared \citep{Langevin2006}. We put forward the hypothesis
that  powerful  katabatic winds (KW) hurtling down existing topographical
slopes up to $\approx$ couple hundreds meters above the surface likely
cause high surface  stresses, lifting the dust that contaminates the
ice. At the same time these winds likely evacuate the dust downstream
in the form of dust clouds lying low in the atmosphere\textbf{ }leading
to an under-estimated optical depth by our method \citep{Doute2013a}.
By the same token we could also explain the fast superficial cleaning
undergone by the \co\  icy deposits at the ground as noted by \citet{Langevin2006}
between $230\lesssim L_{s}\lesssim240\text{\textdegree\ }$. As for
the requirement of significant slopes, we find in the area the mouth
of two chasma and the south polar layered deposits display their most
pronounced scarps down to the surroundings plains. The katabatic mechanism
would be especially efficient for $230\lesssim L_{s}\lesssim240\text{\textdegree\ }$. 

In the next phase, starting at \ls $\approx$ 240\textdegree{}, we
observe a steady rise of seasonal AOD in association with an increasing
day-to-day variability. Conjointly for $244\lesssim L_{s}\lesssim248{^\circ}$
a transition between the ``yellow'' and ``red'' regimes occurs.
That means that, thereafter, the AOD is statistically higher during
the afternoon than during the morning and more so than during the
early hours with low to moderate spread (see also section \ref{sub:Tracking-local-time}
and Table \ref{tab:colors-AOD-LT} for further descriptions). Despite
the growing activity of the second phase, the increase of seasonal
AOD is late compared to unit 2. This is somewhat surprising since
unit 1 is the sector for which the inner crocus line is progressing
the fastest. Our interpretation relies on the fact that, with the
wide separation between the inner and outer crocus lines, spatially
segregated defrosting patterns cover a very extended area for the
considered time period $240\lesssim L_{s}\lesssim260\text{\textdegree\ }$
\citep{Schmidt2009}. As a consequence during daytime, and at the
$\approx$ kilometre scales (small scales), local temperature contrasts
are huge promoting   thermal circulations but limited in their vertical
extent. They become stronger with LT injecting more dust during the
afternoon. Besides during nighttime, because of reduced lateral thermal
contrasts and temperature inversions above high-standing icy surfaces,
the katabatic wind regime could recover for a few hours sweeping the
dust downstream. Conjunction of the two phenomena - katabatic winds
(KW) much more efficiently than small scale thermal circulation (STC)
- restrict atmospheric upward mixing and dust accumulation above the
region to some degree.

By \ls $\approx$ 260\textdegree{} the area corresponding to unit
1 is entirely defrosted. Higher in latitude the separation between
the inner and outer crocus lines is shrinking. Despite the decline
of strong lateral thermal contrasts, we have the emergence of a persistent
regional AOD maximum that grows in value and spatial extent and reaches
its peak at \ls=272-274\textdegree{}. That corresponds to the maximum
level of mean seasonal AOD for the unit which is then also the most
spatially segregated in terms of $\tau_{det}^{k0}$. Unit 1 evolves
from a red regime to a regime for which the night hours display significantly
higher AOD than during daytime even though opacities stay persistently
at high level in agreement with the individual observations (Section
\ref{sub:Transient-events}). These elements forge the following assumption
for the third phase. The conjunction of two mechanisms for atmospheric
dust enhancement in the region could be envisioned. The advection
and upward diffusion (AD2) of dust from the retreating distant edge
of the cap is the first potential mechanism. As explained before preferentially
katabatic winds but also possibly cap edge winds (CW) both able to
carry dust  blow outward from area at higher latitudes that are still
in the second phase.  Thermal gusts and dust devils associated to
the daytime convective turbulence (DCT) \citep{Spiga2011,Spiga2010}
is the second potential mechanism. 

The fourth and last phase corresponds to the decline of seasonal AOD
after \ls $\approx$ 270\textdegree{} with the related diminishing
day-to-day variability or, equivalently, spatial heterogeneity. At
this point the crocus line is remote from the ``cryptic'' region
and has stopped being a source of dust. Hence the thermal gusts and
dust devils which persist exclusively are the best explanation for
moderately efficient dust lifting and upward transport.

\subsection{Unit 2}

For unit 2 we distinguish three successive phases of activity. 

The first one stretches from the beginning of the covered period to
\ls $\approx$ 235-240\textdegree{} when the region is still entirely
covered by the seasonal \co\  deposits. From a starting value of
$\approx$ 0.5 the mean AOD steady grows regardless of the crocus
date (see definition in section \ref{sub:definitions-SSPC}) in many
places which are spatially coherent in their behaviour. Therefore
we suggest that the advection of dust at high altitude (AD1) from
the lower active latitudes is the main origin of the increasing atmospheric
opacity. Meteorological maps (not included here) of wind intensity
and direction were extracted from GCM simulations \citep{Forget1999}
for the southern high latitudes ($\geqslant$60\textdegree{}S) and
examined. They show that poleward air fluxes at synoptic scales are
well developed between 5 and 11 km at this season and nearly reach
the pole. Such fluxes correspond to the return branch of a large scale
convection cell that is established between the cold cap and the surrounding
warming ice free terrains. Due to orographic differences or due to
the different extent of the transition zone, this synoptic feature
could be more strongly established at the longitudes of unit 2 than
at those of unit 1. Indeed the situation differ markedly between the
two units during this early period.

During the second phase $240\lesssim L_{s}\lesssim265{^\circ}$ the
climb of seasonal AOD levels continues at the same overall rate while
the crocus lines sweep the unit. The defrosting area that presents
a geographical mixture of ice-free and still frozen terrains implying
high thermal contrasts is more limited in extent than for unit 1,
of the order of 2 to 5\textdegree{} in latitude. Day-to-day variability
$var_{D2D}$ is also progressively on the rise with greatest values
in the defrosting area. Meanwhile several regional maxima develop
after the passage of the crocus line, and the peak of atmospheric
opacity due to aerosol is reached at \ls $\approx$ 265\textdegree{}.
We put forward the hypothesis that unit 2 shares common processes
with unit 1, those prevailing during the second and third phases of
the latter, with one crucial difference. Similarly to unit 1, we have
in the defrosting area daytime dust lifting associated to small scale
thermal circulation (STC). Contrary to unit 1, where topography allows
for nighttime katabatic winds, the only mechanism to blow efficiently
dust downstream are the daytime thermal cap winds (CW) that could
be established at the mesoscales between the SSPC and the defrosted
terrains. Indeed they likely allow  to some extend vertical diffusion
and transport (AD2) of dust in the atmosphere   towards the lower
latitudes that are entirely defrosted. The whole scheme is compatible
with the examination of the individual OMEGA images that indicate
the occurrence of most transient dust events during the afternoon
in this region. 

In the third phase, when the outer crocus line definitively leaves
unit 2, $var_{D2D}$ declines soon followed by the seasonal level
of AOD. Such a behaviour, equivalent to phase 4 of unit 1, has likely
the same origin: less efficient dust lifting by gusts and dust devils
(DCT).

\subsection{Unit 3}

This unit is characterised by a first phase of gradually enhanced
dust opacity rather similar to the one experienced by unit 2 until
\ls $\approx$ 250\textdegree{}. The difference resides in the fact
that before \ls $\approx$ 245\textdegree{}, the atmospheric dust
layer is more uniform than anywhere else especially in the area of
the ``Mountain of Mitchell''. As a consequence, the day-to-day variability
is also very low and no signs of local time dependencies (at least
for the highest latitudes) can be noted. The advection of dust by
the synoptic circulation (AD1) from active area at lower latitudes
is probably the main origin for the increasing atmospheric opacity
with no contribution of a local activity because all the unit is still
frozen during this phase. 

The situation starts to change after \ls $\approx$ 250\textdegree{}
with the gradual sweeping by the crocus lines, the day-to-day variability
that becomes significant, and the emergence of a ``red'' regime
- dust activity more significant in the afternoon- associated to moderate
to low spread regarding the behaviour of AOD as a function of local
time. In the defrosting area, a relative clearing of the atmosphere
is observed especially since the inner and outer crocus lines are
well separated (\ls $\gtrsim$ 255\textdegree{}). The clearing sometimes
spreads beyond the principal outer crocus line particularly for $266\lesssim L_{s}\lesssim274{^\circ}$
in the area of the ``Mountain of Mitchell'' (then in their last
phase of defrosting) and around Dorsa Argentera around \ls $\approx$
275\textdegree{}. As for the cryptic region in its second phase, the
geographical mixture of ice-free and still frozen terrains should
logically imply local high thermal contrasts in turn promoting intense
but small scale thermal circulation (STC).  The latter would then
be in phase with the diurnal cycle and provide enough surface stresses
to raise the dust \citep{Toigo2002}. However dust clouds are not
detected at significant level by our method nor they are clearly apparent
in the RGB compositions produced from the visible OMEGA images. These
convergent observations strongly support the view of a quiet dust
activity in the transition zone and beyond. Note that, unlike what
goes on in unit 1 during phase 2, the katabatic winds cannot take
over since no significant slopes can be found at the regional scale
in the longitude sector of unit 3. Furthermore the cap winds could
also be inhibited by the broad width of the transition zone, i.e.
the absence of a sharp thermal boundary of regional proportions. The
previous elements converge so as to limit dust opacity around the
SSPC in unit 3 even though it gradually increases.

\subsection{Unit 4}

We think that this unit can be divided into two sub-units each one
being characterised by specific mechanisms despite their very comparable
but likely coincidental behaviour. Sub-unit 4a mainly corresponds
to the lowest latitudes that we cover in the ``anti-cryptic'' sector
whereas sub-unit 4b lies basically in the region of the south permanent
polar cap (SPPC). 

Sub-unit 4a very much behaves like unit 3 but in advance of seasonal
phase. In the period $220\lesssim L_{s}\lesssim240{^\circ}$ advection
of dust at high altitudes from the lower latitudes is the preferred
mechanism for explaining relatively high AOD levels. Around \ls $\approx$
250\textdegree{} with the passage of the crocus line, an apparent
local minimum of seasonal AOD is reached because of the limitations
of our method. Cap edge winds likely blow across the relatively narrow
transition zone lifting the dust but not mixing it high enough for
the AOD to be correctly estimated. Thereafter a progressive and marked
rise of dust opacity intervenes (from 0.55 to 0.8) that we attribute
to the advection of dust by the  cap winds blowing from the defrosting
area that is moving away.  Then the thermal winds fade gradually to
be replaced by  gusts and dust devils of the DCT. 

For sub-unit 4b the initial rise of seasonal AOD until \ls $\approx$
235\textdegree{} and then decline down to $\approx$ 0.5 at \ls $\approx$
250\textdegree{} shows no signs of local time dependencies nor appreciable
low day-to-day variability. This conjunction of factors supports the
hypothesis that dust opacity is then controlled by large scale dynamics
transporting dust high in the polar atmosphere. By \ls $\approx$
250\textdegree{} the atmosphere above the SPPC sees a strong and continuous
enhancement of seasonal AOD. Meanwhile we observe an increasing then
a decreasing day-to-day variability, the latter evolution being well
correlated with the trends affecting the same indicator for the ``cryptic''
region. As far as the local time dependencies are concerned we are
rather in a ``red'' regime (AOD is statistically higher during the
afternoon than during the morning and more so than during the early
hours) with modest spread. Therefore we suggest that the SPPC undergoes
the influence of units 1 and 2 activity by migration of dust clouds
toward the pole at altitudes of a few kilometres thanks to the return
flow of the thermal cap winds (Section \ref{sub:Introducing-circulations}).
Influence of unit 3 and 4a might be more reduced due to the relative
inhibition of the mesoscale thermal winds as evoked earlier.

\section{Summary\label{sec:Summary}}

The OMEGA instrument has acquired a comprehensive set of observations
in the near-infrared (0.93-5.1 microns) at the high southern latitudes
of Mars from mid-winter solstice (\ls=110\textdegree{}, December
2004, MY 27) to the end of the recession (\ls=320\textdegree{}, November
2005). We systematically process a subset of these observations from
\ls =220\textdegree{} to 280\textdegree{} by performing the inversion
of radiative transfer schemes described in \citet{Doute2013a} and
\citet{Vincendon2008} on every top of the atmosphere reflectance
spectra forming the OMEGA images. As a result, we obtained a series
of maps depicting the distribution of aerosol optical depth (AOD)
noted $\tau_{aer}^{k0}$. The maps were normalised in order to correct
for changes due solely to varying atmospheric height because of topography.
They were independently integrated into a common grid generated from
the Hierarchical Equal Area isoLatitude Pixelization of Mars southern
hemisphere. Such an integration, that can be performed at different
spatial resolutions, allows to build the time evolution of the AOD
for each spatial bin of the HEALPix partition. We then separate two
contributions: a mean trend of $\tau_{det}^{k0}$ versus time, i.e.
the baseline, and a highly variable component, i.e. the day-to-day
variability around the baseline with an adapted method. We also implement
a data processing to isolate any dependencies to the local time (diurnal
cycle) of a given bin. The HEALPix system provides means to map the
values ​​of atmospheric opacity or of any derived quantity for a certain
date or averaged over a time interval according to different geographical
systems. Thus we generate time series of orthographic mosaics depicting
spatio-temporal distribution of the seasonal mean values, the variance
and the local time dependence of the AOD. Besides, although the modelling
of the seasonal trend is performed in a pixel-wise manner, the spatial
coherency of the mosaics is excellent. Then we may expect that seasonal
baselines can be gathered, based on similar shapes, into a limited
number of classes. This was confirmed by performing a kmeans classification
that gives four main types of seasonal trends followed by the AOD
$\tau_{det}^{k0}$ in four units segmenting the southern polar region.
Following this complete analysis of the data , each spatio-temporal
unit was studied carefully for searching trends of atmospheric dust
opacity.

As a result a synthetic view (Figure \ref{fig:Synoptic-view}) of
dust activity in the south polar atmosphere in mid spring to early
summer has been established. From this compilation of observations
we propose hypothesis regarding the origin of aerosol activity of
the southern polar region. Different mechanisms are invoked at a variety
of spatial scales in conjunction with the regression stage reached
by the seasonal deposits. In particular we suspect that two mechanisms
might play a major role for lifting and transporting efficiently mineral
particles and create dust events or storms: (i) nighttime katabatic
winds (KW) at locations where a favourable combination of frozen terrains
and topography exists (e.g. unit1, $220\text{\textdegree}\lesssim L_{s}\lesssim255\text{\textdegree}$)
(ii) daytime mesoscale thermal circulation at the edge of the cap,
i.e. cap winds (CW), when the defrosting area (transition zone) is
sufficiently narrow (e.g. unit2, $240\text{\textdegree}\lesssim L_{s}\lesssim260\text{\textdegree}$).
 Indeed this kind of breezes could be inhibited, should the width
of the transition zone be broad, i.e. in the absence of a sharp thermal
boundary of regional proportions (e.g. unit4a, $L_{s}\gtrsim260\text{\textdegree}$).
Thermal circulation at smaller scales (STC) due to the high thermal
contrasts associated to segregated terrains in the transition zone
exist and could also pick up some dust but likely leads to limited
vertical and horizontal transport (e.g. unit3, $250\text{\textdegree}\lesssim L_{s}\lesssim270\text{\textdegree}$).
Far from the seasonal cap, gusts and vortexes associated with Daytime~Convective~Turbulence
(DCT) over ice-free terrains may not be  an efficient mechanism to
inject large amount of dust in the atmosphere explaining the clear
decline of the AOD after \ls $\approx$ 270\textdegree{} for the
whole south polar region. Inside the seasonal cap, advection of dust
at high altitude from the lower active latitudes is likely the main
origin of the increase of the seasonal atmospheric opacity around
\ls $\approx$ 235\textdegree{}, then \ls $\approx$ 275\textdegree{}.

As regards to the production of atmospheric dust in the defrosting
area, the sector with the highest AOD values/variability/increase
spans longitudes 180-300\textdegree{}E (unit 2) around \ls $\approx$
250\textdegree{}. Later (\ls $\approx$ 267\textdegree{}) the situation
has changed drastically and the cryptic sector becomes the most productive
while the longitude sector 300-60\textdegree{}E (unit 4a and 3) remain
moderately dust-generative.

Our work thus contributes to understanding the sequence of phenomena
that control dust activity in the south polar atmosphere and suggests
complex interaction between them. It calls for new simulations of
the martian surface-atmosphere dynamics at mesoscales (between 1000
km and 1 km) considering topographical effects to reproduce the observations
and to confirm the interpretations. In particular taking into account
the gradual defrosting between the inner and outer crocus lines instead
of considering a neat cap edge is of primordial importance. The local
and regional availability of dust and/or sand particles needs to be
addressed as well. Finally the possible interference of the baroclinic
wave system with the convection cell that is established between the
cold cap and the surrounding warming ice free terrains - including
low altitude cap winds and high altitude return flows - must also
be modelled properly.

\section*{Acknowledgements}

We thank the OMEGA team at Institut d'Astrophysique Spatiale for support with sequencing and data downlink activities. This work is
supported by a contract with CNES through its Groupe Système Solaire Program.  We are grateful to X. Ceamanos and A. Spiga for their reading
of the manuscript and for fruitful discussions. We would like to warmly thank both anonymous reviewers for their tremendous work in
reviewing the article and their constructive comments.

{\small \bibliographystyle{elsarticle-harv}

}{\small \par}

\pagestyle{empty}

\end{document}